\RequirePackage{fix-cm}
\documentclass{svjour3}                     
\smartqed  
\usepackage{graphicx}
\usepackage{amssymb}
\usepackage{amsmath}
\usepackage{lscape}
\begin{document}

\title{Secular Dynamics of S-type Planetary Orbits in Binary Star Systems: Applicability Domains of First- and Second-Order Theories
}
\titlerunning{Secular Dynamics of Planets in Binary Star Systems}        

\author{Eduardo Andrade-Ines         \and
        Cristian Beaug\'e \and
				Tatiana Michtchenko \and
				Philippe Robutel
}


\institute{%
        E. Andrade-Ines \at
				Institut de M\'ecanique C\'eleste et de Calcul des \'Eph\'em\'erides (IMCCE) \\
				77 Avenue Denfert Rochereau, Observatoire de Paris, Paris, France \\
			  Instituto de Astronomia, Geof\'{\i}sica e Ci\^{e}ncias Atmosf\'{e}ricas (IAG) \\
				Rua do Mat\~ao, 1226, Universidade de S\~ao Paulo, S\~ao Paulo, Brazil	\\
				\email{eandrade.ines@gmail.com}           			
				\and			
        C. Beaug\'e \at 
				Observatorio Astron\'omico de Cordoba (OAC) \\
				Laprida, 854, Universidad Nacional de C\'ordoba, C\'ordoba, Argentina \\
				\email{beauge@oac.unc.edu.ar}           
				\and
				T. A. Michtchenko\at
        Instituto de Astronomia, Geof\'{\i}sica e Ci\^{e}ncias Atmosf\'{e}ricas (IAG) \\
				Rua do Mat\~ao, 1226, Universidade de S\~ao Paulo, S\~ao Paulo, Brazil	\\
        \email{tatiana.michtchenko@iag.usp.br}    
			  \and
				P. Robutel \at
        Institut de M\'ecanique C\'eleste et de Calcul des \'Eph\'em\'erides (IMCCE) \\
				77 Avenue Denfert Rochereau, Observatoire de Paris, Paris, France \\
				\email{philippe.robutel@obspm.fr}
}

\date{Received: date / Accepted: date}

\maketitle

\begin{abstract}

We analyse the secular dynamics of planets on S-type coplanar orbits in tight binary systems, based on first- and second-order analytical models, and compare their predictions with full N-body simulations. The perturbation parameter adopted for the development of these models depends on the masses of the stars and on the semimajor axis ratio between the planet and the binary.

We show that each model has both advantages and limitations. While the first-order analytical model is algebraically simple and easy to implement, it is only applicable in regions of the parameter space where the perturbations are sufficiently small. The second-order model, although more complex, has a larger range of validity and must be taken into account for dynamical studies of some real exoplanetary systems such as $\gamma$-Cephei and HD 41004A. However, in some extreme cases, neither of these analytical models yields quantitatively correct results, requiring either higher-order theories or direct numerical simulations.

Finally, we determine the limits of applicability of each analytical model in the parameter space of the system, giving an important visual aid to decode which secular theory should be adopted for any given planetary system in a close binary. 

\keywords{Secular Dynamics \and Binary Star \and Second-Order \and Perturbation Theory}

\end{abstract}

\section{Introduction}

Binary stars are frequent in the universe, composing approximately 50\% of main sequence stars (Abt 1979; Duquennoy \& Mayor 1991; Raghavan et al. 2010). Due to inherent difficulties in monitoring the radial velocities of multi-star systems, these have not been primary targets in exoplanet surveys (Eggenberger \& Udry 2010). Still, at least 10\% of the currently known extra-solar planets are hosted in binary stars (Roell et al. 2012).

The gravitational perturbations of a binary star can drastically influence the motion of planetary systems. The dynamical stability of planets in such environments  depends strongly on the orbital and physical parameters of the system (Rabl \& Dvorak 1988; Holman \& Wiegert 1999; Pilat-Lohinger \& Dvorak 2002; Morais \& Giuppone 2012; Andrade-Ines \& Michtchenko 2013). The dynamical effects of the perturbation due to a secondary star can also affect the planetary formation, and even though many studies on this subject have been made (Nelson 2000; Boss 2006; Haghighipour 2006; Nelson \& Kley 2008; Th\'ebault et al. 2009; Giuppone et al. 2011), recent theories still struggle to explain how giant planets can be formed so close to the stability boundary in close binaries (Thebault 2011; Mart\'i \& Beaug\'e 2012, 2015; Silsbee \& Rafikov 2015).

Nevertheless, secular perturbations rule the dynamics of many of these subjects. Giuppone et al. (2011) showed that the planet formation appears more favorable in orbital configurations corresponding to the secular stationary solution. During the later stages of the formation, Michtchenko \& Rodr\'iguez (2011) showed that the migrating planets tend towards stationary configurations, independent of the specific migration mechanism. Andrade-Ines \& Michtchenko (2014) studied the orbital stability of the secular stationary solution. For the particular case of the Habitable Zone of the $\alpha$ Centauri binary system, they also showed that, for orbits close to the secular stationary solution, the variation of the orbital distance to the central star is comparable to that suffered by the Earth despite the strong perturbations of the companion star.

Due to high eccentricities and larger perturbing masses usually found in close binary systems, the classical secular theories based on the Laplace expansion of the disturbing function (\emph{e.g.} Brouwer \& Clemence 1961) are of limited use. An alternative approach is the use of the Legendre expansion of the disturbing function (Heppenheimer 1978; Ford et al. 2000; Georgakarakos 2003, 2005; Laskar \& Bou\'e 2010). Even though this expansion has a larger radius of convergence in terms of the eccentricity, it has a slow convergence rate with respect to the semimajor axis ratio, and is thus usually applied only in hierarchical systems.

Another possible approach is the construction of a semi-analytical model (e.g. Michtchenko \& Malhotra 2004), as applied by Andrade-Ines \& Michtchenko (2014). In this case, the averaging over short-period perturbations is performed numerically over the exact expression of the disturbing function, resulting in a model with no constraints in eccentricities or semimajor axis. However, the procedure is still limited to first-order averaging theories (Giuppone et al. 2011; Andrade-Ines \& Michtchenko 2014). Still in the planetary case, Libert \& Sansottera (2013) developed an analytical second-order in the masses secular model using Laplace coefficients, up to a high degree in the eccentricities. Their model displayed a significant improvement over the first-order model in comparison with the integrations of the exact equations of motion of extra-solar planetary systems, specially close to mean motion resonances. The development, however, was explicitly displayed only for the $\upsilon$ Andromedae system.

A second-order (in the masses) coplanar analytical secular model was developed for the $\gamma$ Cephei binary system, by Giuppone et al. (2011) using a Legendre expansion of the disturbing function. Using Lie-series canonical perturbation theory and considering the restricted case (when the gravitational effects of the planet over the binary stars were neglected), the authors constructed an analytical model that was able to match numerical integrations, at least for initial conditions close to the observed planet. However, the authors also have emphasized that their full model was overly complex and preferred the use of empiric corrections specific for the $\gamma$ Cephei binary system.

Due to the difficulty of constructing and implementing a second-order model, it is of the utmost interest to determine for which orbital configurations the first or second-order models are applicable. The aim of this paper is to develop a general approach for the second-order coplanar secular model and determine the regimes of applicability of the first- and second-order secular models. The limits of applicability will be evaluated in the space of parameters of the problem for planets in S-type orbits (Dvorak 1984), comparing the predictions of each analytical model with direct numerical integrations of the exact equations of motion.

The paper is structured as follows. Section \ref{analytical} presents the analytical foundation of this work, with the expansion of the disturbing function and the application of a canonical perturbation theory for the construction of the first- and second-order secular models. Section \ref{num_sim} presents a numerical method to obtain the main features of the mean secular motion from direct N-body simulations. A comparison between different analytical models and numerical integrations is presented in Section \ref{compare}. Section \ref{app_limits} presents the range of validity of each model in different parametric planes, while applications to real exoplanetary systems are discussed in Section \ref{applications}. Conclusions close the paper in Section \ref{discussions}.

\section{Analytical Models}
\label{analytical}

\subsection{The Disturbing function}
\label{dist_func}

\begin{figure}[th!]
\begin{center}
\includegraphics[width=0.8\textwidth]{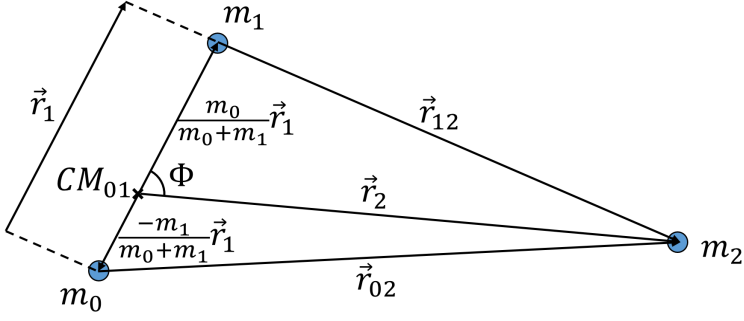}
\caption{The dynamical system under study in the Jacobian reference frame centered at $m_0$. $CM_{01}$ is the center of mass of bodies $m_0$ and $m_1$.} 
\label{jacobi1}
\end{center}
\end{figure}

Let us consider a system composed by a main star of mass $m_0$, a planet of mass $m_1$ and a secondary star of mass $m_2$ in the Jacobian reference frame centered in $m_0$. We denote the position vector of $m_1$ with respect to $m_0$ as $\vec r_1$, while $\vec r_2$ marks the position vector of $m_2$ with respect to the center of mass of $m_0$ and $m_1$ (Figure \ref{jacobi1}). We assume that $|{\vec r_1}| < |\vec r_2|$ for all time. 

The Hamiltonian of the three-body system in these coordinates is given by 
\begin{equation}
{\cal H} = {\cal H}_{K} + {\cal R},
\label{hamiltonian1}
\end{equation}

\noindent where ${\cal H}_K$ is the Keplerian part and ${\cal R}$ is the disturbing function, both given by
\begin{equation}
{\cal H}_K = -\frac{{\cal G}m_0m_1}{2a_1} - \frac{{\cal G}(m_0+m_1)m_2}{2a_2},
\label{hamiltonian2}
\end{equation}
\begin{equation}
{\cal R} = -\frac{{\cal G}m_0m_2}{r_{02}} - \frac{{\cal G}m_1m_2}{r_{12}} + \frac{{\cal G}(m_0+m_1)m_2}{r_2},
\label{perturbadora1}
\end{equation}
In the expressions above, $a_i$ is the Jacobian osculating semimajor axis of the $i$th orbit and  $r_{ij}$ is the distance between the bodies $i$ and $j$, respectively; ${\cal G}$ is the gravitational constant. Terms $1/r_{02}$ and $1/r_{12}$ can be expanded in Legendre polynomials, from which the disturbing function acquires the form
\begin{equation}
{\cal R} = - \frac{{\cal G}}{a_2} \displaystyle\sum_{n=2}^{\infty} {\cal M}_n \alpha^n\frac{\gamma_1^n}{\gamma_2^{n+1}}P_n(\cos\Phi),
\label{perturbadora4}
\end{equation}
\noindent where $P_n(x)$ is the Legendre polynomial of degree $n$, $\Phi$ is the angle between $\vec r_1$ and $\vec r_2$, $\alpha = a_1/a_2$, $\gamma_j = r_j/a_j$ and 
\begin{equation}
{\cal M}_n = m_0m_1m_2\frac{m_0^{n-1} - (-m_1)^{n-1}}{(m_0+m_1)^n}.
\label{massaMl}
\end{equation}
The Legendre polynomials may be written as
\begin{equation}
 P_n(\cos\Phi) = \displaystyle\sum_{q=0}^{n} f_{n,q} \; \textrm{e}^{{\rm i}(2q-n)\Phi},
\label{identidadeWW4}
\end{equation}
(Whittaker \& Watson 1963; Laskar \& Bou\'e 2010) where
\begin{equation}
f_{n,q} = \frac{(2q)!(2n-2q)!}{2^{2n}((n-q)!)^2(q!)^2}.
\label{fnq}
\end{equation}
Substituting (\ref{identidadeWW4}) into (\ref{perturbadora4}) leads to
\begin{equation}
{\cal R} = - \frac{{\cal G}}{a_2} \displaystyle\sum_{n=2}^{\infty} \displaystyle\sum_{q=0}^{n} {\cal M}_n \alpha^n\frac{\gamma_1^n}{\gamma_2^{n+1}} f_{n,q} \textrm{e}^{{\rm i}(2q-n)\Phi}.
\label{perturbadora5}
\end{equation}

In the planar problem, the angle $\Phi$ is given by
\begin{equation}
\Phi = (f_1 - f_2) + \Delta\varpi,
\label{angulos1}
\end{equation}
\noindent where $\Delta\varpi = \varpi_1 - \varpi_2$ and $f_j$ and $\varpi_j$ are the true anomaly and longitude of the pericenter of the $j$th body. The transformation to mean anomalies can be accomplished using Hansen coefficients and Newcomb operators (Plummer 1918; Kaula 1962; Hughes 1981)
\begin{equation}
\gamma_j^n\textrm{e}^{imf_j} = 
\displaystyle\sum_{k_j=-\infty}^{\infty}
X_{k_j}^{n,m}(e_j)
\textrm{e}^{{\rm i}k_jM_j} = 
\displaystyle\sum_{k_j=-\infty}^{\infty}
\displaystyle\sum_{s_j = 0}^{\infty} Y^{n,m}_{s_j+u_{1j},s_j + u_{2j}} e_j^{(2s_j+|m-k_j|)} \textrm{e}^{{\rm i}k_jM_j}
\label{hansen1}
\end{equation}
\noindent where $X_{k_i}^{n,m}$ are the Hansen coefficients, $u_{1j} = \textrm{max}(0,s_j-k_j)$, $u_{2j} = \textrm{max}(0,k_j-s_j)$. Similarly, $Y^{a,b}_{c,d}$ are the Newcomb operators and $M_j$ and $e_j$ are (respectively) the mean anomaly and eccentricity of the $j$th orbit. Introducing (\ref{hansen1}) into (\ref{perturbadora5}) we obtain
\begin{equation}
\begin{array}{rl}
\vspace{0.4cm} 
{\cal R} & = - \displaystyle\frac{{\cal G}}{a_2} \displaystyle\sum_{n=2}^{\infty} \displaystyle\sum_{q=0}^{n} \displaystyle\sum_{k_1=-\infty}^{\infty}\displaystyle\sum_{k_2=-\infty}^{\infty} \displaystyle\sum_{s_1=0}^{\infty} \displaystyle\sum_{s_2=0}^{\infty} {\cal M}_n T^{n,q}_{k_1,k_2,s_1,s_2}\\
 & \times \ \ \ 
\alpha^n
e_1^{2s_1 + |2q-n-k_1|}
e_2^{2s_2 + |n-2q-k_2| }
\textrm{e}^{{\rm i}[(2q-n)\Delta\varpi + k_1M_1 + k_2M_2]},
\end{array}
\label{perturbadora10}
\end{equation}
\noindent where 
\begin{equation}
T^{n,q}_{k_1,k_2,s_1,s_2} = f_{n,q} Y_{s_1+u_A,s_1+u_B}^{n,(2q-n)} Y_{s_2+u_C,s_2+u_D}^{-(n+1),(n-2q)} .
\label{Tnq}
\end{equation}
In this last expression we have denoted $u_A = \textrm{max}(0,k_1 - 2q +n)$, $u_B = \textrm{max}(0,2q - n - k_1)$,  $u_C = \textrm{max}(0,k_2 - n + 2q)$ and $u_D = \textrm{max}(0,n - 2q - k_2)$.

The Newcomb operators can be obtained using recurrence relations (Hughes 1981; Ellis \& Murray 2000), although this calculation can be extremely costly in CPU time. The good news is that $T^{n,q}_{k_1,k_2,s_1,s_2}$ are independent of both initial conditions and the parameters of the system and need only be calculated once. 

We calculated the coefficients $T^{n,q}_{k_1,k_2,s_1,s_2}$ for every value of the set $\{n,q,k_1,k_2,s_1,s_2\}$, truncating the series expansion of the disturbing function considering values of the indexes in the range $2 \leq n \leq 6$, $- 10 \leq (k_1, k_2) \leq 10$ and $0 \leq (s_1, s_2) \leq 8$. We verified that the error caused by this truncation is of the order of the numerical error when comparing the integration of the complete equations of motion of the Hamiltonian to the exact problem for any orbit with $\alpha<0.4$ and $e_1,e_2 < 0.5$. Therefore, this truncation guarantees that any difference between the results of the secular models and the numerical simulations of the exact problem will be only due to the averaging theory adopted in the parameter range $\alpha<0.4$ and $e_1,e_2 < 0.5$, as we discuss in Section \ref{app_limits}.

The advantage of this method is that the factor $T^{n,q}_{k_1,k_2,s_1,s_2}$ is calculated just once and then can be applied to any system with $\alpha<0.4$ and $e_1,e_2 < 0.5$ by just reading a file of $N$ lines. We re-indexed our sum with respect of the line $i$ of that file and we reordered the lines of the file with respect to the magnitude of the term $T^{n,q}_{k_1,k_2,s_1,s_2}\alpha^n e_1^{2s_1 + |2q-n-k_1|} e_2^{2s_2 + |n-2q-k_2|}$, for $\alpha = 0.4$ and $e_1 = e_2 = 0.5$ (for more details see Appendix \ref{coef_num}). This allows us to rewrite the disturbing function as
\begin{equation}
{\cal R} = - \displaystyle\frac{{\cal G}}{a_2} \displaystyle\sum_{i=1}^{N} {\cal M}_{n_i} T_i
\alpha^{n_i}
e_1^{{\rm c}_i}
e_2^{{\rm d}_i}
\textrm{e}^{{\rm i}(p^{(1)}_i M_1 + p^{(2)}_i M_2 + p^{(3)}_i\Delta\varpi)},
\label{perturbadora11}
\end{equation} 

\noindent where we introduced $p^{(1)}_i = k_{1i}$, $p^{(2)}_i = k_{2i}$, $p^{(3)}_i = 2q_i - n_i$,${\rm c}_i =  2s_{1i} + |2q_i-n_i-k_{1i}|$ and ${\rm d}_i=2s_{2i} + |n_i-2q_i-k_{2i}|$ with respect to the new index $i$. The calculated coefficients of the disturbing function are available as an Electronic Supplementary Material, with the files description presented in Appendix \ref{coef_num}.

\subsection{Angle-Action Variables}
\label{angle-action}

To construct our secular model we applied Hori's perturbation theory (Hori 1966; see also Ferraz-Mello 2007) to eliminate the short-period terms associated to the mean anomalies. In the particular case when $a_1 \ll a_2$, the mean motions of both bodies $n_1$ and $n_2$ are of different orders of magnitude and only one of the fast angles has to be eliminated. However, in the general case when $n_2$ is of the same order of $n_1$, we should eliminate both the fast angles. Our set of canonical variables is given by

\begin{equation}
\begin{array}{ll}
M_1, & L_1 = \beta_1\sqrt{a_1\mu_1},\\
M_2, & L_2 = \beta_2\sqrt{a_2\mu_2},\\
\Delta\varpi = \varpi_1 - \varpi_2, & G_1 = L_1\sqrt{1-e_1^2}, \\
\varpi_2, & G_1 + G_2 = L_1\sqrt{1-e_1^2} + L_2\sqrt{1-e_2^2}, \\ 
\end{array}
\label{variables1}
\end{equation}
\noindent where $\beta_i = m_i \sigma_{i-1}$, $\mu_i = {\cal G}\sigma_i$ and $\sigma_i  = \sum_{j=0}^{i}m_j$. At this point, we notice that the disturbing function (\ref{perturbadora11}) does not depend on $\varpi_2$. Therefore its conjugated action $G_1+G_2$ (\emph{i.e.} the total angular momentum) is a constant of motion and the problem can be reduced to three degrees-of-freedom.

From (\ref{hamiltonian1}), (\ref{hamiltonian2}) and (\ref{perturbadora11}), and introducing the definition of $L_i = \beta_i\sqrt{a_i\mu_i}$, the complete Hamiltonian can be written as
\begin{equation}
\begin{array}{rl}
\vspace{0.4cm}
{\cal H} = & -\displaystyle\frac{{\cal G}^2m_0^3m_1^3}{2(m_0+m_1)}\displaystyle\frac{1}{L_1^2} 
- \displaystyle\frac{{\cal G}^2(m_0+m_1)^3m_2^3}{2(m_0+m_1+m_2)} \displaystyle\frac{1}{L_2^2} \\
& - \  {\cal G}^2 \displaystyle\sum_{i=1}^{N} 
T_i {\cal M}_{n_i} {\cal Q}_{n_i} 
L_1^{{\rm a}_i} L_2^{{\rm b}_i}
e_1^{{\rm c}_i}
e_2^{{\rm d}_i} \textrm{e}^{{\rm i}(p^{(1)}_i M_1 + p^{(2)}_i M_2 + p^{(3)}_i\Delta\varpi)},
\end{array}
\label{variables5}
\end{equation}
\noindent where ${\rm a}_i = 2n_i$ and ${\rm b}_i = -2(n_i + 1)$ and
\begin{equation} 
{\cal Q}_{n_i} = \biggr( \frac{m_2}{m_0m_1}\biggl)^{2n_i} \frac{m_2^2(m_0 + m_1)^{3{n_i}+2}}{(m_0+m_1+m_2)^{n_i+1}}.  
\label{Qn}
\end{equation}
Taking the real part of (\ref{variables5}) and introducing the Keplerian terms into the sum as the terms with $i=-1$ and $i=0$ by defining the coefficients
\begin{equation}
\begin{array}{cc}
T_{-1} = 1, & T_{0} = 1 \\ 
n_{-1} = -1 & n_0 = 0 \\
{\cal M}_{-1} =  1 & {\cal M}_{0} = 1  \\ 
{\cal Q}_{-1} =  \displaystyle\frac{m_0^3m_1^3}{2(m_0 + m_1)}& {\cal Q}_{0} =  \displaystyle\frac{m_2^3(m_0+m_1)^3}{2(m_0+m_1+m_2)}   \\ 
{\rm a}_{-1}= -2 & {\rm b}_0 = -2,\\
\end{array}
\label{coeficienteskepler}
\end{equation}
and ${\rm b}_{-1} = {\rm c}_{-1} = {\rm d}_{-1} = p^{(1)}_{-1} = p^{(2)}_{-1} = p^{(3)}_{-1} = {\rm a}_0 = {\rm c}_0 = {\rm d}_0 = p^{(1)}_0 = p^{(2)}_0 = p^{(3)}_0 = 0$, we obtain
\begin{equation}
{\cal H} =
- {\cal G}^2 \displaystyle\sum_{i=-1}^{N} 
T_i {\cal M}_{n_i} {\cal Q}_{n_i} 
L_1^{{\rm a}_i} L_2^{{\rm b}_i}
e_1^{{\rm c}_i}
e_2^{{\rm d}_i} \cos(p^{(1)}_i M_1 + p^{(2)}_i M_2 + p^{(3)}_i\Delta\varpi).
\label{variables6}
\end{equation}

\subsection{Canonical Perturbation Theory}
\label{canonical}

The first step in the application of Hori's method is to reorganize the terms of the Hamiltonian, separating the integrable part (function only of the actions), the secular part (function of the actions and the secular angle $\Delta\varpi$) and the short-period part (function of all variables). We rearranged the terms of the disturbing function such that:

\begin{itemize}
\item for $-1 \leq i \leq N_0$, all the terms have no angular dependence, that is, $p^{(1)}_i = p^{(2)}_i = p^{(3)}_i =0$;
\item for $N_0 < i \leq N_S$, all the terms depend only of the angle $\Delta\varpi$, that is $p^{(1)}_i = p^{(2)}_i = 0$ and $p^{(3)}_i \neq 0$;
\item for $N_S < i \leq N$, all the other terms.
\end{itemize}
Finally, for the small parameter of the problem, we adopted
\begin{equation}
\epsilon = \alpha^2 \frac{m_2}{m_0} = \left( \frac{L_1}{L_2} \right)^4 \frac{m_2^5(m_0 + m_1)^6}{m_0^5m_1^4(m_0+m_1+m_2)^2}.
\label{smallparameter}
\end{equation}

This choice for $\epsilon$ allows the perturbation theory to be applied even to the case where the mass of the perturber $m_2$ is larger than the mass of the central body $m_0$, provided that $\alpha^2$ is small enough. We can therefore formally express the complete Hamiltonian function as

\begin{equation}
{\cal H}(\theta_i,J_i)= {\cal H}_0(J_i) + \epsilon{\cal H}_1(\theta_i,J_i) 
\label{firstorder1}
\end{equation}
\noindent where $\theta_1 = M_1$, $\theta_2 = M_2$ and  $\theta_3 = \Delta\varpi$ are the angle variables, $J_1 = L_1$, $J_2 = L_2$ and $J_3 = G_1$ are their respective conjugated actions and
\begin{equation}
{\cal H}_0 = - \  {\cal G}^2 \displaystyle\sum_{i=-1}^{N_0} T_i {\cal M}_{n_i} {\cal Q}_{n_i} L_1^{{\rm a}_i} L_2^{{\rm b}_i}e_1^{{\rm c}_i}e_2^{{\rm d}_i}, 
\label{firstorder2}
\end{equation}
\begin{equation}
\begin{array}{rl}
\vspace{0.2cm}
{\cal H}_1 = & - \  {\cal G}^2 \displaystyle\sum_{i=N_0+1}^{N_S} T_i {\cal M}_{n_i} {\cal K}_{n_i} L_1^{{\rm a}_i-4} L_2^{{\rm b}_i+4}e_1^{{\rm c}_i}e_2^{{\rm d}_i} \cos(p^{(3)}_i\Delta\varpi) - \\
& \  - {\cal G}^2 \displaystyle\sum_{i=N_S+1}^{N} T_i {\cal M}_i {\cal K}_i L_1^{{\rm a}_i-4} L_2^{{\rm b}_i+4}e_1^{{\rm c}_i}e_2^{{\rm d}_i}  \cos(p^{(1)}_i M_1 + p^{(2)}_iM_2 + p^{(3)}_i\Delta\varpi),
\end{array}
\label{firstorder3}
\end{equation}
\noindent where $e_1$ and $e_2$ are functions of the variables $L_1$, $L_2$ and $G_1$ and of the constant of motion $G_1+G_2$, and we introduced a new mass factor 
\begin{equation}
{\cal K}_i = \frac{m_2^{2i-3}}{m_0^{2i-5}m_1^{2i-4}}\frac{(m_0+m_1)^{3i-4}}{(m_0+m_1+m_2)^{i+1}}.
\label{firstorder4}
\end{equation}

To construct our second-order secular theory, our goal is to find a Lie-type transformation $E_{B^*}$ of the variables $(\theta_i,J_i)$, generated by the function

\begin{equation}
B^*(\theta_i^*,J_i^*) = \epsilon B_1^*(\theta_i^*,J_i^*) + \epsilon^2 B_2^*(\theta_i^*,J_i^*) + \mathcal{O}(\epsilon^3)
\label{firstorder4.1}
\end{equation}
to a new set of variables $(\theta_i^*,J_i^*)$, such that the new Hamiltonian

\begin{equation}
\begin{array}{rl}
\vspace{0.4cm}
E_{B^*} {\cal H}(\theta_i^*,J_i^*) = & {\cal H}^*(\theta_3^*,J_i^*) \\
= & {\cal H}^*_0(J_i^*) + \epsilon H_1^*(\theta_3^*,J_i^*)+\epsilon^2 {\cal H}^*_2(\theta_3^*,J_i^*) + \varrho(\theta_3^*,J_i^*,\epsilon^3)
\label{firstorder4.2}
\end{array}
\end{equation}

\noindent is independent of the angles $M_1^*$ and $M_2^*$, and $\varrho$ is the remainder of order $\mathcal{O}(\epsilon^3)$.

Recalling that the Hamiltonian is time independent, expanding the Lie series on the left-hand side of (\ref{firstorder4.2}) and identifying the terms in same order in $\epsilon$, we get

\begin{equation}
\begin{array}{rl}
\vspace{0.4cm}
{\cal H}_0^* = & {\cal H}_0, \\
\vspace{0.4cm}
{\cal H}_1^* = & {\cal H}_1  + \left\{ {\cal H}_0,B_1^* \right\}, \\
\vspace{0.4cm}
{\cal H}_2^* = & {\cal H}_2  +  \displaystyle\frac{1}{2} \left\{{\cal H}_1+{\cal H}_1^*,B_1^* \right\} + \left\{ {\cal H}_0,B_2^* \right\}. \\
\end{array}
\label{firstorder4.3}
\end{equation}

From Eq. \ref{firstorder4.3} we identify the \emph{homological equation}

\begin{equation}
\nu_1\frac{\partial B_k^*}{\partial M_1^*} +\nu_2\frac{\partial B_k^*}{\partial M_2^*}+\nu_3\frac{\partial B_k^*}{\partial \Delta \varpi^*}= \Psi_k - {\cal H}^*_k , 
\label{secondorder13}
\end{equation}

\noindent where $\nu_i$ are the three frequencies of non-resonant coplanar problem, defined by $\nu_i = \partial {\cal H}^*_0/\partial J_i^*$ (see Appendix \ref{ap01}), and $\Psi_k$ is a known function once all the previous $(k-1)$ normalization steps are performed (for more details, see Ferraz-Mello 2007, chapter 6). However, in (\ref{secondorder13}) both functions $B_k^*$ and $H_k^*$ are unknown. This indetermination is solved, without loss of generality, by adopting the averaging rule:

\begin{equation}
{\cal H}_k^* = \langle \Psi_k \rangle_{M_1^*,M_2^*}.
\label{secondorder17}
\end{equation}

From (\ref{firstorder3}) and (\ref{secondorder17}), for $k=1$, we obtain the first-order solution:
\begin{equation}
{\cal H}_1^* = - \  {\cal G}^2 \displaystyle\sum_{i=N_0+1}^{N_S} T_i {\cal M}_{n_i} {\cal K}_{n_i} L_1^{*{\rm a}_i-4} L_2^{^*{\rm b}_i+4}e_1^{*{\rm c}_i}e_2^{*{\rm d}_i} \cos(p^{(3)}_i\Delta\varpi^*).
\label{firstorder18}
\end{equation}

\noindent where $e_1^*$ and $e_2^*$ are expressed as functions of $G_1^*$ and of the parameters $L_1^*$, $L_2^*$ and $G_1^* + G_2^*$. From this point forward, we will refer to the model composed by ${\cal H}_0^* + \epsilon{\cal H}_1^*$ as the \emph{First-order secular model}. From (\ref{secondorder13}) and (\ref{firstorder18}) the first-order of the generating function can be calculated and yields
\begin{equation}
B_1^* =  - {\cal G}^2 \displaystyle\sum_{l=N_S+1}^{N} \displaystyle\frac{T_l {\cal M}_{n_l} {\cal K}_{n_l} L_1^{*{\rm a}_l-4} L_2^{*{\rm b}_l-4}e_1^{*{\rm c}_l}e_2^{*{\rm d}_l}}{p^{(1)}_l\nu_1 + p^{(2)}_l\nu_2 + p^{(3)}_l\nu_3} \sin(p^{(1)}_l M_1^*+ p^{(2)}_l M_2^*+ p^{(3)}_l \Delta\varpi^*).
\label{secondorder19}
\end{equation}
 
Introducing (\ref{firstorder18}) and (\ref{secondorder19}) into (\ref{secondorder13}) and applying the averaging rule (\ref{secondorder17}) for $k=2$, after a long but straightforward calculation, we finally obtain the second-order solution, which, explicitly, is written as:
\begin{equation}
\begin{array}{rl}
\vspace{0.4cm}
{\cal H}_{2}^* = & \displaystyle\frac{{\cal G}^4}{4}\Biggl\{  \displaystyle\sum_{i=N_0+1}^{N} \displaystyle\sum_{j=N_S+1}^{N} \displaystyle\frac{\sigma_i \delta_{p^{(1)}_j,-p^{(1)}_i} \delta_{p^{(2)}_j,-p^{(2)}_i} T_i {\cal M}_{n_i} {\cal K}_{n_i}T_j {\cal M}_{n_j} {\cal K}_{n_j}}{-p^{(1)}_i\nu_1 - p^{(2)}_i\nu_2 + p^{(3)}_j\nu_3} \\
\vspace{0.4cm}
& \times \  L_1^{*{\rm a}_i + {\rm a}_j-8} L_2^{*{\rm b}_i+ {\rm b}_j-8}e_1^{*{\rm c}_i+{\rm c}_j}e_2^{*{\rm d}_i+{\rm d}_j} \\
\vspace{0.4cm}
& \times \Biggl[ \ \ \  p^{(1)}_i \biggl(\displaystyle\frac{{\rm a}_j + {\rm a}_i - {\rm c}_j- {\rm c}_i}{L_1^*} + \displaystyle\frac{{\rm c}_j+{\rm c}_i}{L_1^*e_1^{*2}}\biggr) - p^{(2)}_i \biggl(\displaystyle\frac{{\rm b}_j + {\rm b}_i - {\rm d}_j - {\rm d}_i}{L_2^*} + \displaystyle\frac{{\rm d}_j + {\rm d}_i}{L_2^*e_2^{*2}} \biggr) \\ 
\vspace{0.4cm}
& \ \ \ + \ p^{(3)}_i\biggl(\displaystyle\frac{{\rm d}_j\sqrt{1-e_2^{*2}}}{L_2^*e_2^{*2}} - \displaystyle\frac{{\rm c}_j\sqrt{1-e_1^{*2}}}{L_1^*e_1^{*2}} \biggr) - \ p^{(3)}_j\biggl(\displaystyle\frac{{\rm d}_i\sqrt{1-e_2^{*2}}}{L_2^*e_2^{*2}} - \displaystyle\frac{{\rm c}_i\sqrt{1-e_1^{*2}}}{L_1^*e_1^{*2}} \biggr) \Biggr] \\ 
\vspace{0.4cm}
&  \ \ \times \cos[(p^{(3)}_i+p^{(3)}_j)\Delta\varpi^*] \Biggr\} \\ 
\vspace{0.4cm}
& + \ \displaystyle\frac{{\cal G}^4}{4}\Biggl\{  \displaystyle\sum_{i=N_0+1}^{N} \displaystyle\sum_{j=N_S+1}^{N} \displaystyle\frac{\sigma_i \delta_{p^{(1)}_j,p^{(1)}_i} \delta_{p^{(2)}_j,p^{(2)}_i} T_i {\cal M}_{n_i} {\cal K}_{n_i}T_j {\cal M}_{n_j} {\cal K}_{n_j}}{p^{(1)}_i\nu_1 + p^{(2)}_i\nu_2 + p^{(3)}_j\nu_3} \\
\vspace{0.4cm}
& \times  \ L_1^{*{\rm a}_i + {\rm a}_j-8} L_2^{*{\rm b}_i+ {\rm b}_j-8}e_1^{*{\rm c}_i+{\rm c}_j}e_2^{*{\rm d}_i+{\rm d}_j} \\
\vspace{0.4cm}
& \times \Biggl[ \  - \  p^{(1)}_i \biggl(\displaystyle\frac{{\rm a}_j-{\rm a}_j - {\rm c}_i + {\rm c}_j}{L_1^*} + \displaystyle\frac{{\rm c}_j - {\rm c}_i}{L_1^*e_1^{*2}}\biggr) - \  p^{(2)}_i \biggl(\displaystyle\frac{{\rm b}_j-{\rm d}_j - {\rm b}_i + {\rm d}_j}{L_2^*} + \displaystyle\frac{{\rm d}_j - {\rm d}_i}{L_2^*e_2^{*2}}\biggr) \\ 
\vspace{0.4cm}
& \ \ \ - \ p^{(3)}_i\biggl(\displaystyle\frac{{\rm d}_j\sqrt{1-e_2^{*2}}}{L_2^*e_2^{*2}} - \displaystyle\frac{{\rm c}_j\sqrt{1-e_1^{*2}}}{L_1^*e_1^{*2}} \biggr) - \ p^{(3)}_j\biggl(\displaystyle\frac{{\rm d}_i\sqrt{1-e_2^{*2}}}{L_2^*e_2^{*2}} - \displaystyle\frac{{\rm c}_i\sqrt{1-e_1^{*2}}}{L_1^*e_1^{*2}} \biggr) \Biggr] \\ 
\vspace{0.4cm}
& \ \ \times \cos[(p^{(3)}_i-p^{(3)}_j)\Delta\varpi^*] \Biggr\}, \\ 
\end{array}
\label{secondorder28}
\end{equation}
\noindent where $e_1^*$ and $e_2^*$ are expressed as functions of $G_1^*$ and of the parameters $L_1^*$, $L_2^*$ and $G_1^* + G_2^*$ and we introduced $\sigma_l$, defined as

\begin{equation}
\sigma_l = \left\{ 
\begin{array}{l}
0, \textrm{ for } l \leq N_0 \textrm{;} \\
2, \textrm{ for } N_0 < l \leq N_S\textrm{;} \\ 
1, \textrm{ for } l> N_S\textrm{;}
\end{array}
\right.
\label{def_sigma}
\end{equation}

\noindent and $\delta_{i,j}$ is the Kronecker delta, defined as

\begin{equation}
\delta_{i,j} = \left\{ 
\begin{array}{l} 
1, \textrm{ for } i=j; \\
0, \textrm{ for } i\neq j;
\end{array}
\right.
\label{def_delta}
\end{equation}

It is worthy emphasizing that the second-order solution presented above is only valid when we consider the non-resonant condition $i\nu_1 + j\nu_2 + k\nu_3 \neq 0$, with $(i,j,k) \in \mathbb{Z}^3 \{(0,0,0)\}$. Finally, the complete secular Hamiltonian up to second order in $\epsilon$ is given by 

\begin{equation}
{\cal H}^* = {\cal H}_0^* + \epsilon{\cal H}_1^* + \epsilon^2{\cal H}_2^*,
\label{Hsecondorder}
\end{equation}

\noindent with ${\cal H}_0^*$, ${\cal H}_1^*$ and ${\cal H}_2^*$ given by (\ref{firstorder2}), (\ref{firstorder18}) and (\ref{secondorder28}), respectively. From this point forward, we will refer to this model as the \emph{second-order secular model}, and the equations of motion of the one degree-of-freedom system are given by

\begin{equation}
\begin{array}{lr}
\displaystyle\frac{\partial {\cal H}^*}{\partial \Delta\varpi^*} = -\displaystyle\frac{dG_1^*}{dt} \;\;\; , & \;\;\; \displaystyle\frac{\partial {\cal H}^*}{\partial G_1^*} = \displaystyle\frac{d\Delta\varpi^*}{dt},
\end{array}
\label{secondorder29}
\end{equation}

\noindent and, $L_1^*$, $L_2^*$, and $G_1^* + G_2^*$ are constants of motion, since their conjugated angle variables ($M_1^*$, $M_2^*$ and $\varpi_2^*$, respectively) do not appear explicitly in the secular Hamiltonian. Let us remark that the secular Hamiltonians possessing only one degree-of-freedom are integrable. As a consequence, chaotic motions cannot be produced by these models. The evolution of the orbital parameters of the secular problem can be obtained by simultaneously integrating Eqs. (\ref{secondorder29}) numerically.

\subsection{Representation of the Secular Motion Using the Classical Model of Heppenheimer (1978)}
\label{hepp78}

Although the model developed above was constructed for the general three-body problem, most secular models assume that $m_1 \ll m_0,m_2$. In the limit of the restricted three-body problem ($m_1 \to 0$), bodies $m_0$ and $m_2$ move in fixed ellipses as described by the two-body problem. Up to first-order in the masses, it is possible to obtain an expression of the disturbing function which is exact with respect to $e_2$ (\emph{e.g.}, Kaula 1962; Laskar \& Bou\'e 2010). Limiting the expansion in Legendre polynomials to $P_2$ (quadrupole problem) and truncating the perturbation to order $\mathcal{O}(e_1^2)$, Heppenheimer (1978) obtained the averaged disturbing function in orbital elements as
\begin{equation}
{\cal R}_{\textrm{Hep}} = \frac{{\cal G} m_2}{(1-e_2^2)^{3/2}}\frac{a_1^2}{a_2^3}\left[ \frac{1}{4} + \frac{3}{8}e_1^2 - \frac{15}{16}\frac{a_1}{a_2}\frac{e_1 e_2}{(1-e_2^2)}\cos(\Delta\varpi) \right],
\label{hepp1}
\end{equation}
\noindent where we omitted the constant terms. Introducing the non-singular variables
\begin{equation}
\begin{array}{c}
h = e_1\sin(\Delta\varpi),\\
k = e_1\cos(\Delta\varpi),
\end{array}
\label{hepp2}
\end{equation}
the modified Lagrange-Laplace planetary equations will be, up to order $\mathcal{O}(e_1^2)$ (Brouwer \& Clemence 1962):
\begin{equation}
\begin{array}{cc}
\displaystyle\frac{d h}{dt} = g_s(k - e_{1F})\;\;\; , & \;\;\; \displaystyle\frac{d k}{dt} = - g_sh,\\
\end{array}
\label{hepp5}
\end{equation}
\noindent where
\begin{equation}
g_s = \frac{3}{4}\frac{m_2}{m_0}\frac{a_1^3}{a_2^3} \frac{n_1}{(1-e_2^2)^{3/2}}
\label{hepp6}
\end{equation}
is the forced secular frequency and 
\begin{equation}
e_{1F} = \frac{5}{4} \frac{a_1}{a_2} \frac{e_2}{(1 - e_2^2)}
\label{hepp7}
\end{equation}
is the forced eccentricity. In both equations $n_1 = \sqrt{{\cal G} m_0/a_1^3}$ is the mean-motion of the planet. The general solution of the system of Eqs. (\ref{hepp5}) acquire the form
\begin{equation}
k(t) = e_{1p} \cos(g_s t + \phi_0) + e_{1F},
\label{hepp8}
\end{equation}
\begin{equation}
h(t) = e_{1p} \sin(g_s t + \phi_0),
\label{hepp9}
\end{equation}
\noindent where $e_{1p}$ (proper eccentricity) and $\phi_0$ (phase angle) are constants of integration determined by the initial conditions. We can see from Eqs. (\ref{hepp6}) and (\ref{hepp7}) that $g_s$ and $e_{1F}$ are functions only of the parameters of the problem. According to Eqs. (\ref{hepp8}) and (\ref{hepp9}), the secular orbits define circles in the $k,h$ plane centered in $k=e_{1F}$ and with only a single frequency $g_s$. Both are independent of the initial conditions of the planetary orbit.  

The trajectory starting with $e_{1p} = 0$ gives $h(t) = 0$ and $k(t) = e_{1F}$, and therefore $e_1(t) = e_{1F}$, is a stationary solution or \emph{fixed point}. Since there is only one fixed point, located in the semi-plane $k>0$, we can conclude that the secular angle $\Delta\varpi$ will either circulate or oscillate around 0. The resulting oscillation around $\Delta\varpi=0$ is also known as {\it Mode I} (Michtchenko \& Ferraz-Mello 2001).

\subsection{Extension to Higher-Order Theories}
\label{sec_ef}

The model of Heppenheimer (1978) is a good approximation to the problem if $a_1/a_2$ and $e_1$ are sufficiently small. For larger values of these quantities, the expressions presented above for both $g_s$ and $e_{1F}$ no longer yield quantitatively accurate values, although the topology of the secular problem remains unaltered (Giuppone et al. 2011). 

Analytical approximations for the solutions including second-order terms have so far been estimated either by empirical approximations (\emph{e.g.} Th\'ebault et al. 2006) or by functional approximations (Giuppone et al. 2011). In both cases, however, the resulting expressions for $g_s$ and $e_{1F}$ are not general and valid only for a sub-set of the parameter space. In particular, the expressions for $g_s$ found by these authors are only valid for $\gamma$-Cephei, but fail for other values of the perturbing mass and eccentricity. 

In order to obtain better estimations, we can use the first- and second-order secular models presented in Section \ref{canonical}. However, since the expressions for the averaged Hamiltonian are too complex, we must open hand of explicit close formulas and determine $g_s$ and $e_{1F}$ numerically. This can be achieved employing the geometric method introduced by (Michtchenko \& Malhotra 2004) and later applied by Michtchenko et al. (2006) and Andrade-Ines \& Michtchenko (2014). This method consists in finding the eccentricity $e_{1F}$ that gives the extreme value of the Hamiltonian for $\Delta\varpi = 0$ and given values of the parameters $L_1$, $L_2$ and $G_1 + G_2$ (the reader is referred to Michtchenko \& Malhotra 2004 for a detailed description). 

In particular, the forced eccentricity is the solution of the algebraic equation
\begin{equation}
\frac{\partial {\cal H}}{\partial G_1} \Big|_{G_{1F},\Delta\varpi=0} = 0,
\label{modogs3}
\end{equation}
\noindent where $G_{1F} = L_1\sqrt{1-e_{1F}^2}$. Similarly, the secular frequency for the first- and second-order models are given by
\begin{equation}
g_s = \frac{\partial }{\partial G_1}\langle{\cal H}\rangle_{\Delta\varpi}, 
\label{gsec1}
\end{equation}
\noindent where $G_{1} = L_1\sqrt{1-e_1^{2}}$. The analytical expressions of the secular frequencies for the first- and second-order models are presented at the Appendix \ref{ap01}.

\section{Numerical Simulations}
\label{num_sim}

Since any analytical theory is expected to be valid only for certain initial conditions or values of the parameters of the system, it is important to be able to deduce the main features of the secular solutions using numerical integration of Newton's equations of motion. In this section we show how to determine the families of forced eccentricities and secular frequencies from N-body simulations.

\subsection{Numerical Determination of the Forced Eccentricity and Secular Frequency}
\label{num_det}

In the secular problem, the fixed point at $e_1 = e_{1F}$ and $\Delta \varpi=0$ is a stationary orbit with zero secular amplitude. When the averaged variables are transformed back to osculating values, the resulting trajectory will be a quasi-periodic solution with 3 main frequencies $\nu_1 = n_1$, $\nu_2= n_2$ and $\nu_3 = g_s$, although with a zero amplitude associated with the secular frequency $\nu_3 = g_s$. Therefore, a fixed point in the secular (averaged) problem will correspond to a quasi-periodic orbit in osculating elements with frequencies $n_1$ and $n_2$.

Determining such a quasi-periodic solution from N-body simulations can be a challenging task. Fortunately, there exist several numerical tools that can be employed to simplify this work. One was used by Noyelles et al. (2008) and later by Couetdic et al. (2010) based on frequency analysis of the numerical integration (Laskar 1990; Michtchenko et al. 2002). This method has proved to be very efficient and yields accurate results. Its main steps are summarized as follows:

\begin{enumerate}
\item Numerical integration of an orbit for a given set of initial conditions;
\item Harmonic decomposition of the time series of the orbital elements to determine the fundamental frequencies;
\item Quasi-periodic decomposition of the time series in function of the fundamental frequencies;
\item Elimination of the terms depending on $\nu_3$ and construction of a new time series of the orbital elements;
\item Determination of a new set of initial conditions from the time series of the orbital elements with $\nu_3$ suppressed.  
\end{enumerate}

This process is iterative in nature, with each new set of initial conditions being closer to the solution and the convergence is generally fast, reducing the amplitudes of the secular components by 2 orders of magnitude in just 4 steps (Couetdic et al. 2010).

In order to identify all the 3 frequencies of motion with the harmonic decomposition, the integrations must be long enough to include at least one secular period. Moreover, the integration step must be small enough such that the keplerian period of the planet is identifiable. For the present work, we used the \emph{NAFF} (Laskar 1999) algorithm for the harmonic decomposition, with a time step of $\Delta t = 2\pi/(3n_1)$ and a total time of integration of $T = 12\pi/g_s$, where an approximate value for $g_s$ was adopted following (\ref{hepp6}). Since we expect the real secular frequency to be different from that estimated from first-order models, we used a total integration time at least 6 times the approximate secular period.

The iterative process was stopped whenever the relative difference between the initial eccentricities of the planet in two consecutive iterations was smaller than $0.1\%$. Once the initial conditions of the quasi-periodic orbit were determined, the forced eccentricity was estimated by
\begin{equation}
e_{1F} = \langle e_1(t) \cos (\Delta\varpi(t)) \rangle_T, 
\label{eFnum}
\end{equation}

\noindent with the secular frequency $\nu_3$ obtained at the quasi-periodic decomposition step of the last iteration of the method.

\subsection{Non-Convergent Cases}
\label{RMM-prog}

The method described above provides an accurate approximation of the fundamental frequencies as long as the trajectory satisfies two conditions: (i) is regular (i.e. not chaotic), and (ii) is not dominated by mean-motion resonances (MMRs). If one of these conditions is met, then the quasi-periodic secular approximation is no longer valid and the iterative process will not be convergent. Even though these cases are not covered by the secular models, the analysis of these orbits is important in order to compare the predictions of the analytical models with N-body simulations.

Applying the quasi-periodic decomposition method in an unstable or a resonant orbit will lead to an inaccurate determination of the fundamental frequencies that will compromise the convergence of the method. For this reason, it was imposed in the algorithm that if the convergence condition was not satisfied in 20 steps, the orbit would go through a stability check. 

The stability of the orbits was numerically estimated by determining the proper mean motion of the planet $n_1$ with a quasi-periodic decomposition routine (Robutel \& Laskar 2001): the analysis of the first two thirds of the data defined a value  $n_{1a}$ for the mean-motion, while the last two thirds of the data was used to calculate a second value $n_{1b}$. If the difference $|n_{1a} - n_{1b}|/n_{1a}$ was found to be greater than $10^{-3}$, the orbit was considered unstable and the algorithm issued fictitious values of $e_F = 1$ and $g_s = 0$.

\section{Accuracy of Different Analytical Models}
\label{compare}

To assess the quality of the analytical secular models, in this section we compare the results obtained from our first- and second-order models (Section \ref{canonical}), with the classical model of Heppenheimer (1978) (Section \ref{hepp78}) and with numerical simulations (Section \ref{num_det}), which we will take as the exact solution. As a working example we chose the binary star system \emph{HD 196885 AB}; physical and orbital data of this system, as well as the data of the detected planet around the star A, are presented in Table \ref{dadoshd196885}. 

\begin{table}
\caption{Physical and orbital parameters of the System HD 196885 AB (Chauvin et al., 2011).}
\label{dadoshd196885}
\begin{tabular}{llllllll}
\hline\noalign{\smallskip}
Body & $a$ (au) & $e$ & $i$ (deg) & $M$ (deg) & $\omega$ (deg) & $\Omega$ (deg) & $m$ $(M_\odot)$ \\ 
\noalign{\smallskip}\hline\noalign{\smallskip}
A & - &-  &-  &-  &-  &-  & 1.3\\  
B & 21 & 0.42 & 116.8 & 121  & 241.9 & 79.8 & 0.45\\  
b & 2.6 & 0.48 & 116.8* & 349  & 93.2 & 79.8* or 259.8*& $3.186 \times 10^{-3}$* \\  
\noalign{\smallskip}\hline
\end{tabular}
{\scriptsize *Most stable/probable solution according to Giuppone et al. (2012).}
\end{table}

\subsection{Forced Eccentricity}
\label{forced_ecc}

\begin{figure}[!ht]
\begin{center}
\includegraphics[width=0.9\textwidth]{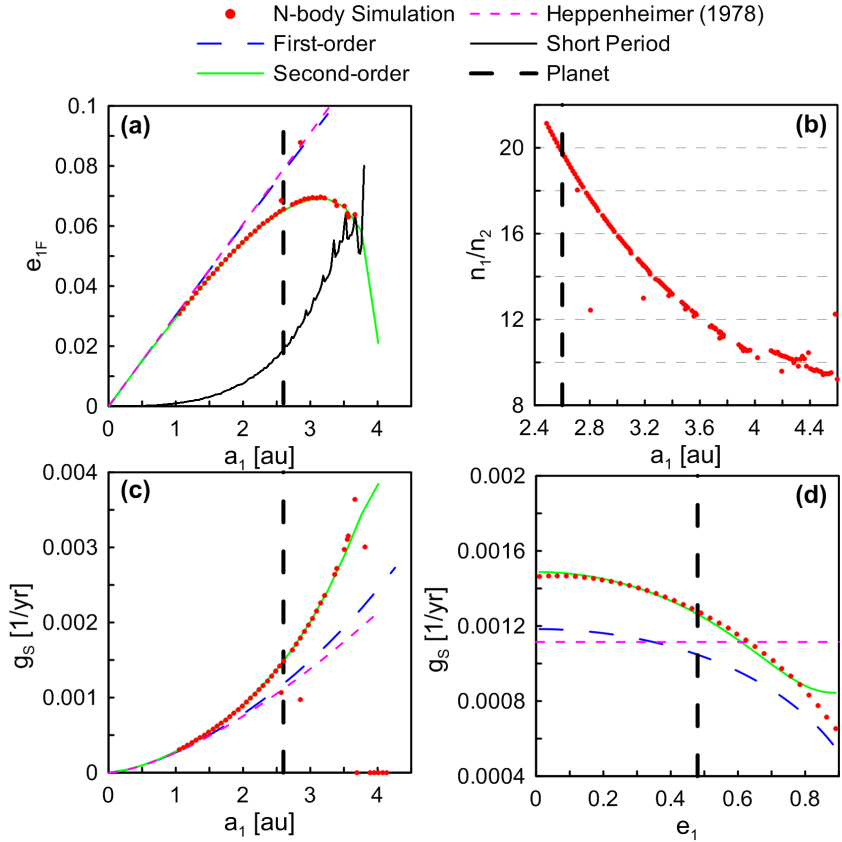}
\caption{{\bf (a)} Secular forced eccentricity $e_{1F}$ as function of the semimajor axis $a_1$, calculated with different analytical models, compared with the results of numerical simulations (red dots). the black curve shows the amplitude of the short-period variations. {\bf (b)} Averaged mean-motion ratio $n_1/n_2$, as function of the initial osculating semimajor axis $a_1$, calculated from N-body simulations. The location of several first-degree MMRs are marked with horizontal lines. {\bf (c)} Secular frequency $g_s$, as function of the semimajor axis $a_1$, calculated for different models, as well as with the numerical integration (red dots). {\bf (d)} 
Secular frequency as function of the eccentricity $e_1$ of the perturber for $a_1 = 2.6$ au. In all panels the values of $a_1$ and $e_1$ of the planet are marked by dashed vertical lines. The scattered red dots are non-convergent solutions obtained by the numerical method (see Section \ref{RMM-prog}).}
\label{ex-curvas}
\end{center}
\end{figure}

Figure \ref{ex-curvas} (a) shows the family of stationary secular solutions (i.e. forced eccentricity $e_{1F}$) as function of the semimajor axis of the planet $a_1$, for the system \emph{HD 196885 AB}. All other parameters of the system were fixed according to the values given by Table \ref{dadoshd196885}. The red circles correspond to the result obtained from the numerical integrations, the magenta curve shows the solution using the model by Heppenheimer (1978), while  the blue and green curves present the solutions obtained from our first- and second-order models, respectively. The dashed vertical line marks the present osculating semimajor axis of the detected planet ($a_1 = 2.6$ au).

Both first-order models show a linear dependence of the forced eccentricity with the semimajor axis, while the second-order model and the numeric solution show a significant quadratic component. As expected, for sufficiently small values of $a_1$ all models coincide, while increasingly large deviations are seen for orbits closer to the perturber. At the present location of the planet the predictions of the first-order model are not quantitatively correct, indicating that any model for the secular dynamics of this system should include second-order terms. 

In the same frame, the black curve represents the amplitude of the short period oscillations, calculated as the difference between the maximum and minimum values that $e_1$ reached in a single keplerian period of the star. This amplitude shows a strong correlation with the difference in forced eccentricity between both the first- and second-order models. This is not surprising, since the magnitude of the second-order terms scales with the short-period variations (see Eq. \ref{secondorder19}). 

\subsection{Mean-Motion Resonances}

Fig. \ref{ex-curvas} {\bf (b)} shows the dependence of the numerically determined mean motion ratio $n_1/n_2$ with the semimajor axis of the planet, close to the family of secular stationary solutions. 

As discussed in Section \ref{RMM-prog}, the crossing of MMRs, defined by the condition $i n_1 + j n_2 \approx 0, \ i,j \in \mathbb{Z}$, can lead to instabilities that can hinder the convergence of the iterative method described at Section \ref{num_det}. As a result, due to the MMRs, we see ``gaps'' in the curve of Fig. \ref{ex-curvas} {\bf (b)}. We have found that the gaps appear for each $n_1/n_2 \approx i, \ 9 \leq i \leq 18 \in \mathbb{Z} $, with the gaps getting larger with the decrease of $i$, up until $n_1/n_2 \approx 9 $, when we have the stability limit for this system.

Even though the resonant problem is a complex subject and each MMR should be studied individually, we can still estimate empirically where in the phase space the MMRs may begin to play an important part in the dynamical evolution of this system. For instance, we identify the first significant resonance as the gap with the lowest semimajor axis, that appears at the 18:1 MMR, at $a_1 \approx 2.8$ au. We see that the planet is located very close to the 20:1 MMR, but the short time dynamical effects of this resonance were not detected by this method and therefore we conclude that the secular dynamics will still play the major part in the dynamical evolution of this system.

\subsection{Secular Frequencies}
\label{num-secular_frequencies}

Figure \ref{ex-curvas}(c) shows the variation of the secular frequency $g_s$ as function of $a_1$. As before, all other parameters were taken from Table \ref{dadoshd196885}. In both panels we present the first- and second-order models (blue and green curves, respectively), the Heppenheimer (1978) model (magenta curves) and the solution obtained from the exact equations of motion (red circles).
 
As before, the second-order solution presents an excellent agreement with the numerical results, while the first-order models predict smaller values of the secular frequency for initial conditions closer to the perturber. Also, it is interesting to note that our first-order version now shows a noticeable (albeit small) deviation with respect to Heppenheimer's version, which was not evident in the case of the forced eccentricity. The scatter of the numerical results for larger semimajor axis is due to the effect of mean-motion resonances. In particular, the 10/1 MMR, located at $a_1 \simeq 4$ au caused non-convergence of the numerical method for many initial conditions, assigning to them an artificial value $g_s=0$. Other mean-motion resonances are also noticeable, although with smaller effect.

\begin{figure}[!ht]
\begin{center}
\includegraphics[width=0.9\textwidth]{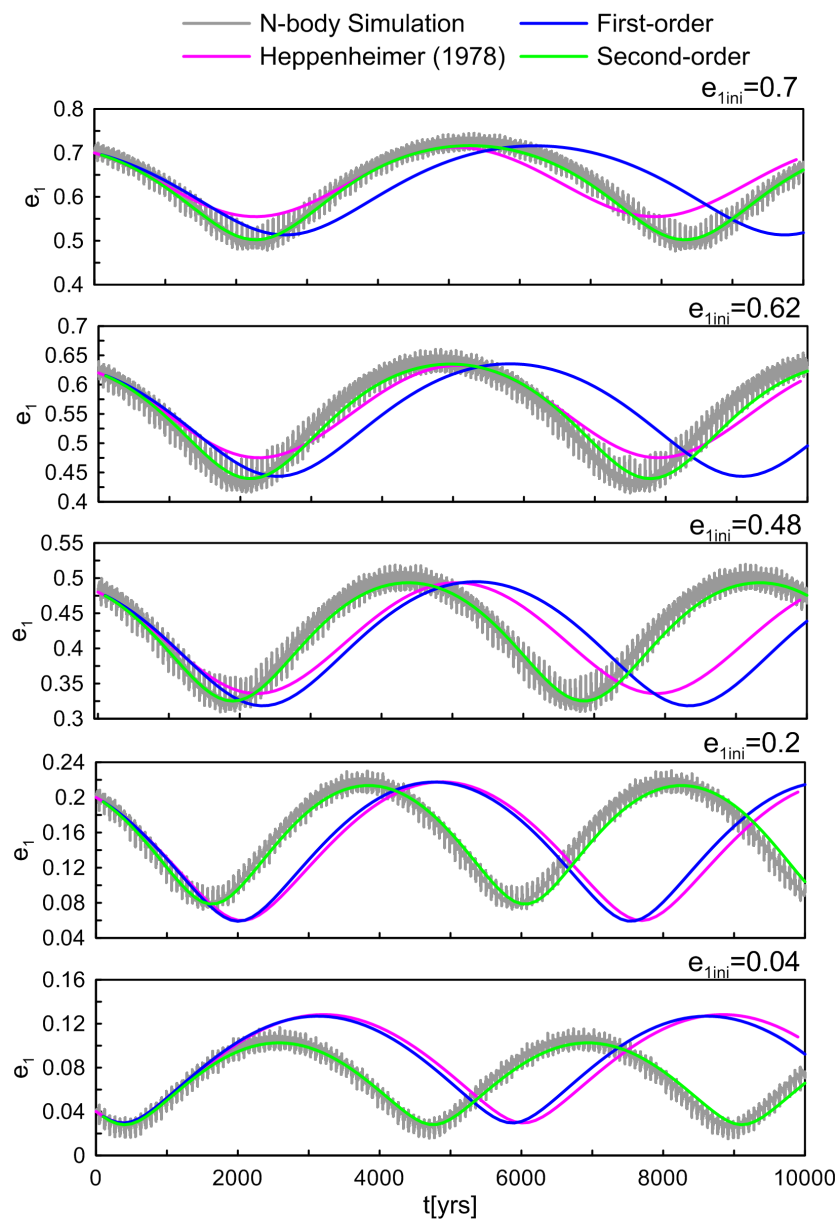}
\caption{Time evolution $e_1$ for the planet around HD 196885 considering different initial values (indicated on top of each frame). Results obtained from N-body simulations are shown in gray, while predictions of different analytical models are indicated in color curves. Note that Heppenheimer's model shows a good fit for the secular frequency for $e_{\rm ini} = 0.62$ in accordance with the lower-right panel at Figure \ref{ex-curvas}.}
\label{ex-orbita}
\end{center}
\end{figure}

These results are similar to those found by Giuppone et al. (2012) in the case of $\gamma$-Cephei, indicating that a second-order secular theory may be not only desirable but actually necessary in many planetary systems around close binary stars.

While the analytical models of Heppenheimer (1978) and Giuppone et al. (2012) assumed zero-amplitude secular solutions at the fixed point, our model has the advantage of allowing to map finite amplitude oscillations and find the complete secular solutions of the system even if the initial conditions are far from the stationary value. These will occur whenever the initial value of the eccentricity $e_1$ is different from the forced value $e_{1F}$ and/or $\Delta \varpi \ne 0$.

One of the consequences of finite-amplitude oscillations is that the secular frequency is different from that given by its stationary value. Figure \ref{ex-curvas} (d) shows the dependence of $g_s$ with the initial eccentricity of the planet. It has a maximum value at $e_1 = e_{1F}$, and decreases for increasing amplitudes of oscillation. Our second-order model shows a very good agreement with the full numerical simulations up to $e_1 \approx 0.8$, a value higher than expected due to the truncation of the disturbing function for $e_1 < 0.5$. 

In contrast, the secular frequency predicted by Heppenheimer's model shows no dependence with $e_1$. Therefore, there should be always a value of $e_1$ for which the secular frequency determined from both the second-order and Heppenheimer's models coincide. Particularly for the system HD 196885, with $a_1 = 2.6$ au, this happens for $e_1 \approx 0.6$, which is close to the current value of the eccentricity of the planet (see the dashed vertical line in Fig. \ref{ex-curvas}(c)). We emphasize, however, that this is a coincidence and there is no way of predicting with just first-order models for which value of $e_1$ this will happen for different systems.

To illustrate the dependency of $g_s$ with $e_1$, Figure \ref{ex-orbita} shows (in gray) the result of five numerical integrations which differ only in the initial values of the eccentricity. The predictions of the different analytical models are depicted in colored lines. In all cases our second-order model shows a very good agreement with the N-body results, not only with respect with the frequency but also in the amplitude of oscillation. None of the other models appear reliable, although, again, Heppenheimer's solution does show a good fit for the frequency for $e_{1ini} = 0.62$.

\section{Applicability Limits}
\label{app_limits}

Although the example described above shows that a second-order secular model must be employed in some real planetary systems around binary stars, others are not so extreme and may be adequately mapped with a simple first-order model. Since the second-order theory is, by construction, much more complex, it is important to predict when it is really necessary and when it may be avoided. Similarly, even a second-order model will breakdown for initial conditions too close to the perturber, and it is also important to have some idea of its range of validity. 

In this section we present a graphical representation of the {\it Limits of Applicability} of each analytical secular model, in terms of the main parameters of the system: mass and orbit of the secondary star, and semimajor axis of the planet. As proxy we adopt the forced eccentricity determined by each model, as compared with the value obtained from direct numerical integrations.

\subsection{Definition of the Limits}
\label{sec-def_lim}

\begin{figure}[t!]
\begin{center}
\includegraphics[width=0.8\textwidth]{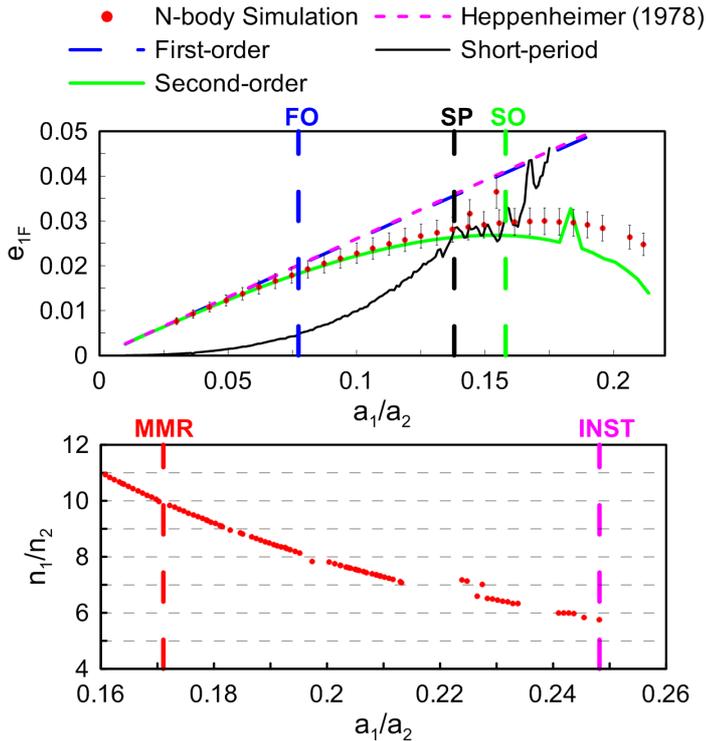}
\caption{Families of stationary secular solutions for a fictitious binary system with $m_0 = m_2 = 1 M_\odot$, $m_1 = 10^{-4} M_\odot$, $a_2 = 1$ au, $e_2 = 0.2$ and $\Delta\varpi =0$. Top panel:  forced eccentricity as function of the semimajor axis ratio determined from numerical simulations (red circles). The error bars correspond to a relative error of $10\%$.
Color curves show the predictions of different analytical models: first- and second-order models (blue and green, respectively) and Heppenheimer (1978) model (magenta). Numerical estimation of the amplitude of short-period oscillations are indicated in black. The vertical dashed lines represent the characteristic limits {\bf FO} (blue), {\bf SO} (green) and {\bf SP} (black); see text for details. Bottom panel: Mean-motion ratio as function of the semimajor axis ratio. The vertical dashed lines represent the {\bf MMR} (red) and {\bf INST} (magenta) characteristic limits. The  {\bf MAcD} limit(Andrade-Ines \& Michtchenko 2014) occurs for larger semimajor axis and is not drawn in this plot.}
\label{def_lim}
\end{center}
\end{figure}

The top panel of Figure \ref{def_lim} shows the variation of $e_{1F}$ and the amplitudes of short-period oscillations as function of the initial semimajor axis, for a fictitious system with parameters $m_0 = m_2 = 1 M_\odot$, $m_1 = 10^{-4} M_\odot$, $a_2 = 1$ au and $e_2 = 0.2$. In the case of the forced eccentricity, numerical results are again shown in red circles, while colored curves indicate the predictions of different analytical models. We will denote by:
\begin{equation}
\Delta e_{1F} = \frac{|{e_{1F}}_{\rm model}-{e_{1F}}_{\rm exact}|}{{e_{1F}}_{\rm exact}}
\end{equation}
as the relative error of the forced eccentricity estimated by a given model, with respect to its exact (numerically determined) value. The amplitude of the short-period variations are shown with a black curve and were determined numerically with an N-body simulation. Finally, the bottom panel shows the mean-motion ratio $n_1/n_2$ as function of the semimajor axis ratio $a_1/a_2$. 

The vertical dashed lines in both graphs represent a series of \emph{characteristic limits}, defined as:

\begin{itemize}

\item {\bf FO} : Value of $a_1/a_2$ where $\Delta e_{1F} = 0.1$, calculated with the 1st-order model;

\item {\bf SO} : Value of $a_1/a_2$ where $\Delta e_{1F} = 0.1$, calculated with the 2nd-order model;

\item {\bf MMR} : Lowest value of $a_1/a_2$ for which mean-motion resonances cause significant non-convergence of the secular models;

\item {\bf SP} : Value of $a_1/a_2$ where the amplitude of the short-period oscillations equals the forced eccentricity;

\item {\bf INST} : Lower limit of $a_1/a_2$ leading to orbital stability. Beyond this point some (but not all) initial conditions result in collision or in an expulsion of the planet from the binary system.

\item {\bf MAcD} : Upper limit of $a_1/a_2$ leading to orbital stability. Beyond this point {\it all} initial conditions result in collision or in an expulsion of the planet from the binary system.

\end{itemize}

The {\bf SP} limit is an estimative of the region where the short-period dynamics may play an important role, and where their amplitude rivals that of the secular dynamics. As discussed in Section \ref{forced_ecc}, the generating function $B_1^*$ (Eq. \ref{secondorder19}) depends only on the short-period terms; consequently, the larger the amplitude of these terms, the higher the order of the averaging theory we may need to apply. Therefore, the applicability limits {\bf FO} and {\bf SO} should be correlated with the {\bf SP} limits.

The lower instability limit {\bf INST} signals the appearance of resonance overlap where some (but not all) initial conditions exhibit unstable motion. Full orbital instability (for all initial conditions) roughly corresponds to the limit {\bf MAcD}, which was estimated following the criterion developed in Andrade-Ines \& Michtchenko (2014), adopting $e_1=e_{1F}$. According to this model, global instability is said to occur for all values of the semimajor axis satisfying the condition:
\begin{equation}
a_1(1+e_{1F}) \ge R_{{cr}}, 
\label{SPSeq1}
\end{equation}
\noindent where 
\begin{equation}
\begin{array}{rl}
R_{cr}/a_2\approx & 0.66823 - 0.63740e_2 - 0.74549 (m_2/m_0) + 0.45496e_2 (m_2/m_0) \\ 
& + 1.0492 (m_2/m_0)^2 - 0.23179 e_2 (m_2/m_0)^2 - 0.87722 (m_2/m_0)^3 \\
& + 0.31541 (m_2/m_0)^4, \\
\end{array}
\label{approximation1}
\end{equation}
for $0.1 \leq m_2/m_0 \leq 1.0$, and 
\begin{equation}
\begin{array}{rl}
R_{cr}/a_2\approx & 0.45265 - 0.41921e_2 - 0.070754(m_2/m_0) +0.039617e_2 (m_2/m_0) \\ 
& + 0.010865(m_2/m_0)^2 - 2.1394\times10^{-3}e_2 (m_2/m_0)^2 \\ 
& -9.3729\times 10^{-4}(m_2/m_0)^3 + 3.3886\times 10^{-5}(m_2/m_0)^4,\\
\end{array}
\label{approximation2}
\end{equation}
for $1.0 < m_2/m_0 \leq 10.0$. 

Note that we assume that the initial conditions of the planet coincide with the stationary secular solution, which is not necessarily the case (e.g. HD196885). However, it is sufficient for most purposes and serves as a proxy for the stability limit for S-type orbits in binary systems.

\subsection{Parametric Planes}
\label{parametric_planes}

We calculated the characteristic limits defined in Section \ref{sec-def_lim} for fictitious binary systems with a central mass of $m_0 = 1 M_\odot$ and seven different values for the secondary mass:
$m_2/M_\odot = (0.1,0.2,0.5,1,2,5,10)$. These values were chosen to include cases in which the planet orbits the most massive component, as well as situations in which the opposite occurs. The semimajor axis of the binary was set at $a_2 = 1$ au, and the eccentricity was again varied, this time taking values $e_2 = (0.1,0.2,0.3,0.4,0.5,0.6)$. The mass of the planet was chosen equal to $m_1 = 10^{-4} M_\odot$ (roughly, twice the mass of Neptune) and its initial semimajor axis varied in the interval $a_1 \in [0.01,0.6]$ au.

\begin{figure}[t!]
\begin{center}
\includegraphics[width=0.85\textwidth]{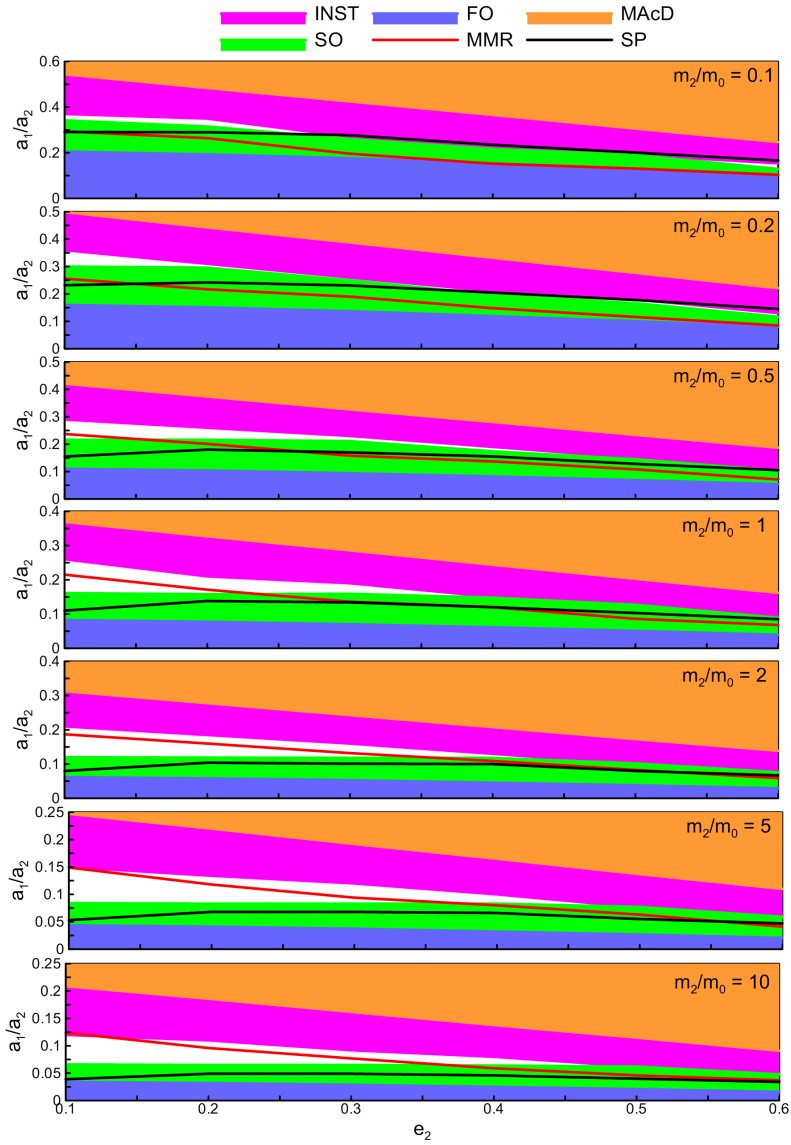}
\caption{Limits of applicability of first-order secular models (area in blue) and second-order models (area in green) in the $(e_2,a_1/a_2)$ parametric plane, for seven different values of $m_2/m_0$. The region in white denotes initial conditions where the relative error of the second-order model surpasses $10\%$, but still with no detected influence of mean motion resonances. The black curve marks the semimajor axis where the amplitude of short-period terms is equal to $e_{1F}$ (i.e. {\bf SP}), while the lower limit of semimajor axis where significant mean-motion resonances were detected are shown by the red curve. The region in magenta correspond to initial conditions found to be dynamically unstable (i.e. {\bf INST}), while the limit predicted by the MAcD-criterion are colored orange.} 
\label{mcte1}
\end{center}
\end{figure}

\begin{figure}[t!]
\begin{center}
\includegraphics[width=0.85\textwidth]{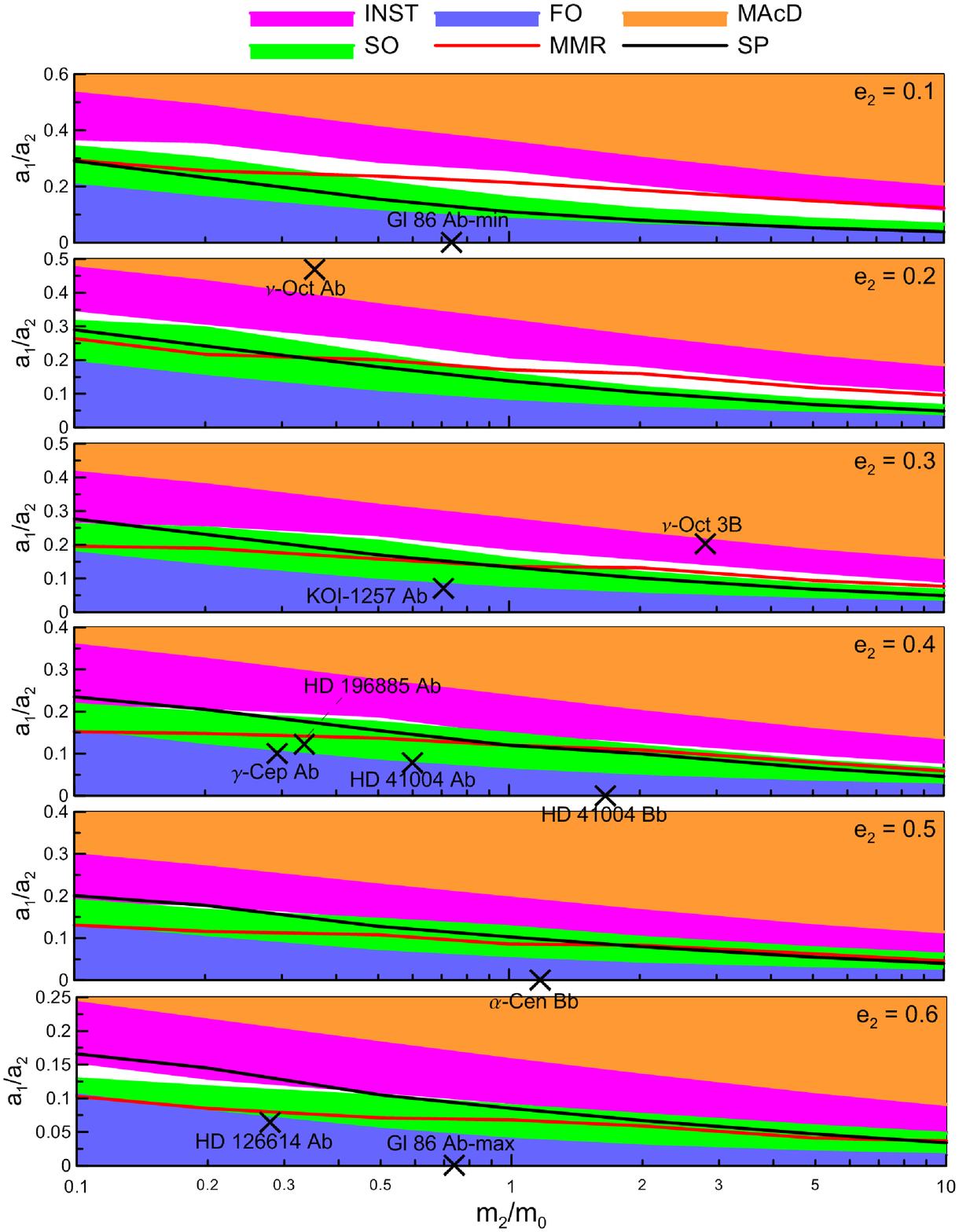}
\caption{Limits of applicability of first-order secular models (area in blue) and second-order models (area in green) in the $(m_2/m_0,a_1/a_2)$ parametric plane, for six different values of $e_2$. The color code is the same as in the previous Figure. The black crosses represent the location of planets in real binary systems, as summarized in Table \ref{dadosplanetas}.} 
\label{ecte1}
\end{center}
\end{figure}

Results are summarized in Figures \ref{mcte1} and \ref{ecte1}. The first shows seven parametric plots in the $(e_2,a_1/a_2)$ plane, constructed for different values of the mass ratio $m_2/m_0$. 
Figure \ref{ecte1} shows six parametric plots in the $(m_2/m_0,a_1/a_2)$ plane, each for a different value of the eccentricity $e_2$. The definition of each colored region is specified in the captions.

As expected, the applicability of the first-order model decreases with larger values of $m_2$ and $e_2$, but even in the most favorable cases it never exceeds $a_1/a_2 \simeq 0.2$, reaching values below $a_1/a_2 \simeq 0.1$ for massive binary companions in eccentric orbits. Even so, as indicated by the black crosses in Figure \ref{ecte1}, the secular dynamics of a few known planetary systems may be well described by this simple analytical model. The second-order model has a larger region of applicability, reaching approximately twice the range in $a_1/a_2$ for a given binary companion.

Figure \ref{mcte1} shows that, for a fixed value for the mass ratio, the influence of $e_2$ on the limits of applicability of the analytical secular models is mostly due to the stability boundary (area in magenta). For example, we can see at the bottom plane, with $m_2/m_0 = 10$, that the applicability limits are approximately constant as function of $e_2$. More details of this feature will be discussed in Section \ref{families}.

Figure \ref{ecte1} shows that the white region is larger for lower values of $e_2$. For the case $e_2 = 0.1$, we can see a white region even for the mass ratio $m_2/m_0 = 0.1$, and, as the mass ratio gets larger, the white region also gets wider. 

Both Figs. \ref{ecte1} and \ref{mcte1} show that the only case that there is a white region above the curve in black is for the case $e_2 = 0.6$. However, we must remember that the disturbing function was developed to describe precisely systems with $e_i \leq 0.5$ (Section \ref{dist_func}). Therefore, this difference may be due to the inadequate development of the disturbing function for this case specifically, even though we can see that the models are applicable to most of the space of parameters. 

For the cases where $e_2 \leq 0.5$ we can see that the regions in white are always located above the curves in black, which indicates that, in fact, these are regions strongly influenced by the short-periodic terms. As it was discussed in Section \ref{sec_ef}, this is a strong evidence that the regions in white should be able to be properly described by extending the perturbation theory to third or higher order.

We can see as well that there is a large portion of the green regions that are located above the curves in red. This indicates that the width of the first detected MMRs are narrow enough such that the dynamics in their neighborhood is still strongly secular. At this point, it is also worthy emphasizing that the stability limits presented in both Figs. \ref{ecte1} and \ref{mcte1} were calculated for coplanar systems with $e_1$ close to the stationary solution and for the initial values of the angular variables at 0. All these regions were located above the curves in red, therefore MMRs could play a meaningful part in the system dynamics such as to form stability islands for higher values of eccentricity in the space of parameters. Andrade-Ines \& Michtchenko (2014) shows as well that there is a very high dependence to the planetary initial eccentricity and inclination to the plane of the binaries in the system dynamics and stability.

\subsection{Families of Secular Stationary Solutions}
\label{families}

\begin{figure}[!ht]
\begin{center}
\includegraphics[width=0.98\textwidth]{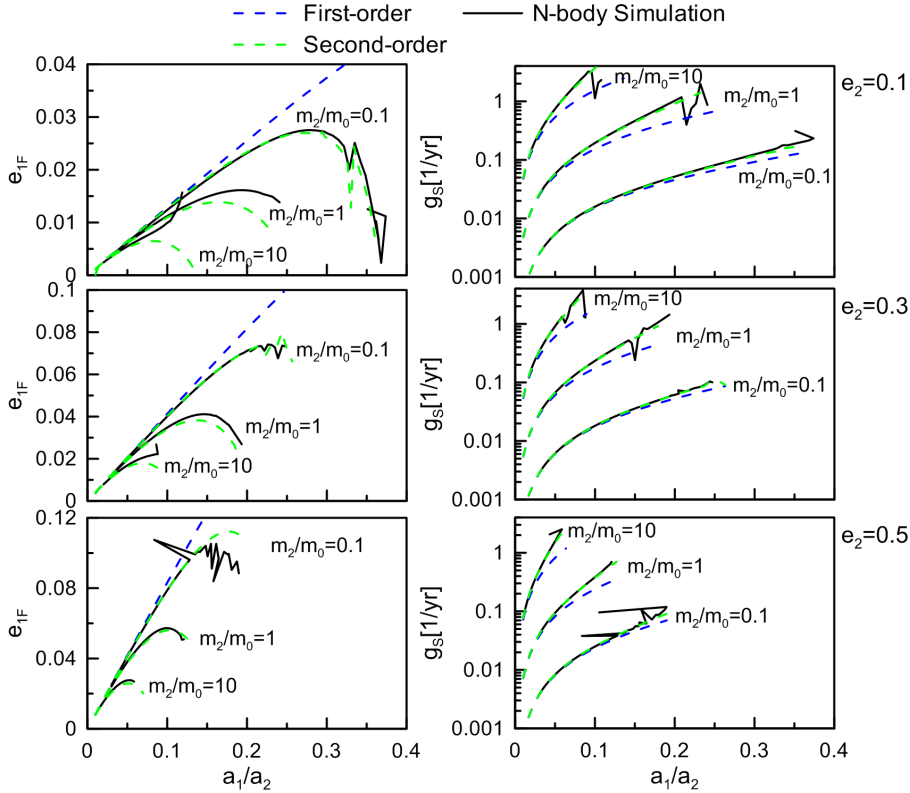}
\caption{Forced eccentricities (left panels) and secular frequencies (right panels) of stationary secular solutions calculated with the first- and second-order models (blue and green curves, respectively), as well as values obtained from N-body simulations (black curves). The top panels show the results obtained for $e_2 = 0.1$, the middle panels for $e_2 = 0.3$ and the bottom panels for $e_2 = 0.5$.}
\label{ex_familias}
\end{center}
\end{figure}

Figure \ref{ex_familias} shows the families of forced eccentricities (left panels) and secular frequencies (right panels) for zero-amplitude secular solutions as function of the semimajor axis ratio $a_1/a_2$. These were calculated for $m_0 = 1 M_\odot$, $m_1 = 10^{-4} M_\odot$, $a_2 = 1$ au, $\Delta\varpi =0$ and three different values of $e_2$. Within each plane we also varied the mass ratio between the secondary and central star. 
 
For the forced eccentricities, the second-order secular model yields very accurate results in the case $m_2/m_0 = 0.1$, even for large values of $e_2$ and $a_1/a_2$. Thus, this theory appears very reliable in systems orbiting the larger of the binary components. For equal mass binaries ($m_2/m_0 = 1$), the precision of the analytical model is restricted to smaller values of the semimajor axis, especially for binaries in more circular orbits. Finally, less accurate results are obtained for $m_2/m_0 = 10$, indicating that our theory has limited applicability in systems where the planet orbits the less massive star. 

For the secular frequencies (right-hand panels), the second-order model presents a good approximation to the exact solution in all cases, even in the case of planets orbiting the less massive star. The first-order frequencies, however, are only reliable for small values of semimajor axis. The difference in values of $g_s$ between both models can reach up to a factor two, depending on the parameters of the system. 

It is also worth mentioning that, while the general behavior of $g_s$ of the different models is scaled by the semimajor axis ratio, the values of the secular frequencies will depend on the individual values of $a_1$ and $a_2$. The secular frequencies will also depend on the initial value of the planetary eccentricity $e_1$, as was discussed in Section \ref{num-secular_frequencies}.

\section{Application to Detected Planetary Systems}
\label{applications}

The black crosses in Figure \ref{ecte1} show the location of several exoplanetary systems of interest in the parametric plane. These include both confirmed and unconfirmed planetary candidates orbiting one component in a tight binary system (orbital separation between stars $a_2 < 50$ au). The orbital and physical parameters of these systems are shown in Table \ref{dadosplanetas}. For a quantitative comparison between the efficiency of the different models, we present in Table \ref{dadosplanetasmodelo} the forced eccentricity $e_F$ and the secular frequency $g_S$ calculated with the second-order secular model, as well as the error in percentage one would have by adopting the first-order model for each system.

\begin{table}
\caption{Physical and orbital parameters of planets and candidates in S-type orbits with $a_2 <50$ au.}
\label{dadosplanetas}
\begin{tabular}{llllllllll}
\hline\noalign{\smallskip}
System &  $m_0$ &  $m_1 \sin(i)$ & $m_2$ & $a_1$ & $a_2$ &  $e_1$ & $e_2$ & Ref \\ 
 & $[M_\odot]$ &  $[10^{-3}M_\odot]$ & $[M_\odot]$ & [au] & [au] &  &  &\\
\noalign{\smallskip}\hline\noalign{\smallskip}
$\nu$ Oct Ab$^a$     & 1.4 & $2.386$ & 0.5 & 1.2 & 2.55 & 0.123 & 0.2359 & (1)\\
$\nu$ Oct Triple$^a$ & 0.496 & $42 $ & 1.4 & 0.524 & 2.565 & 0.67 & 0.2504 & (2) \\
KOI-1257$^b$        & 0.99 & $1.384 $ & 0.7 & 0.382 & 5.3 & 0.772 & 0.31 & (3) \\
HD 41004 Ab       & $0.700$ & $2.424 $ & $0.420$  & 1.60  & 20.0 & 0.48 & 0.40 & (4), (5)\\
HD 41004 Bb       & $0.420$ & $17.53$ & $0.700$  & 0.0177  & 20.0 & 0.081 & 0.40 & (4), (5), (6)\\
$\gamma$ Ceph Ab  & $1.40$ & $1.765 $ & $0.410$  & 2.05 & 20.2 & 0.05 & 0.41 & (7), (8), (9)\\
HD 196885 Ab      & $1.33$ & $2.815$ & $0.450$  & 2.60  & 21.0 & \textbf{0.48} & 0.42 & (10)\\
$\alpha$ Cen Bb$^c$   & $0.934$ & $0.003435$ & $1.10$  & 0.04 & 23.4 & 0.0 & 0.518 & (11), (12)\\
Gl 86 Ab min $^d$ & 0.8 & $3.827 $ & 0.59  & 0.11 & 30.58 & 0.046 & 0.1 & (13), (14)\\
Gl 86 Ab max $^d$ & 0.8 & $3.827 $ & 0.59  & 0.11 & 69.8 & 0.046 & 0.61 & (13), (14)\\
HD 126614 Ab      & $1.15$  & $0.3626$ & $0.320$  & 2.35 & 36.2 & 0.30 & $\leq 0.6$ & (15)\\
\noalign{\smallskip}\hline
\end{tabular}
{\scriptsize Notes: 
$^a$ Unconfirmed due to orbital instability of coplanar solution. Planet could lie on a highly inclined or retrograde orbit;
$^b$ Planet is a candidate;
$^c$ The detection of this planet is contested by Hatzes (2013);
$^d$ Parameters of the binary are not well known, constrained by the relation $a_2(1+e_2) \approx 28$ au. 
References:
(1) Ramm et al. (2009),
(2) Morais \& Correia (2012),
(3) Santerne et al. (2014),
(4) Zucker et al. (2004),
(5) Roell et al. (2012),
(6) Santos et al. (2002),
(7) Neuh{\"a}user et al. (2007),
(8) Endl et al. (2011),
(9) Reffert \& Quirrenbach (2011),
(10) Chauvin et al. (2011),
(11) Pourbaix (1999),
(12) Dumusque et al. (2012),
(13) Queloz et al. (2000),
(14) Farihi et al. (2013),
(15) Howard et al. (2010).}
\end{table}

\begin{table}
\caption{Forced eccentricities $e_F$ and secular frequencies $g_S$ calculated for the systems presented on Table \ref{dadosplanetas} with the second-order secular models. The secular frequencies were calculated for a fictitious planet at the stationary secular solution ($e_1=e_F$)  and for the actual value of the planetary eccentricity given by Table \ref{dadosplanetas}. We present as well the difference in percentage $\sigma$ of the first-order solution relatively to the second-order one. The $\nu$ Oct Ab$^a$ and $\nu$ Oct Triple$^a$ are located in above the MMR curve in Fig. \ref{ecte1} and therefore the secular models are not applicable.}
\label{dadosplanetasmodelo}
\begin{tabular}{llllllllll}
\hline\noalign{\smallskip}
System   & $e_{F_ {\rm SO}}$ & $\sigma (e_F)$ & $g_{S_{\rm SO}}(e_F)$ & $\sigma (g_S(e_F))$ & $g_{S_{\rm SO}}(e_1)$ & $\sigma (g_S(e_1))$ \\
   & $[10^{-2}]$ & \% & [$10^{-4}$/yr] & \% & [$10^{-4}$/yr] & \% \\
\noalign{\smallskip}\hline\noalign{\smallskip}
$\alpha$ Cen Bb$^c$   & 0.1512 & 0.0370 & 0.05356 & 0.0907 & 0.05356 & 0.0907 \\
Gl 86 Ab min $^d$  & 0.04530 & 0.265 & 0.04019 & 0.412 & 0.04015 & 0.518 \\
Gl 86 Ab max $^d$  & 0.188 & 1.84 & 0.0067 & 0.554 & 0.006674 & 0.660 \\
HD 41004 Bb        & 0.05367 & 1.85 & 0.020 & 1.67 & 0.0195 & 1.62 \\
HD 126614 Ab       & 7.090 & 7.30 & 2.344 & 11.0 & 2.23023 & 6.47 \\
KOI-1257$^b$         & 2.854 & 8.26 & 74.9 & 13.9 & 70.46 & 8.46 \\
$\gamma$ Ceph Ab   & 5.645 & 10.8 & 9.05 & 15.3 & 9.057 & 15.3 \\
HD 41004 Ab        & 4.26 & 11.8 & 9.27 & 16.0 & 7.872 & 1.08 \\
HD 196885 Ab       & 6.56 & 20.4 & 14.6179 & 23.7 & 12.45 & 10.4 \\
\noalign{\smallskip}\hline
\end{tabular}
\end{table}

In the following we analyse each system individually:

\begin{itemize}
\item $\nu$ Oct Ab - Located in the orange region, indicating strong orbital instability. This system has been the subject of discussion in many works (Eberle \& Cuntz 2010; Quarles et al. 2012; Go{\'z}dziewski et al.(2013); among others), suggesting that the planet may orbit the central star in a retrograde orbit. 

\item $\nu$ Oct Triple -  An alternative description of the same system, proposed by Morais \& Correia (2012), composed of a binary sub-system instead of a single secondary star. The planet predicted in this scenario is located in the magenta region, with strong dynamical effects from mean-motion resonances;

\item  KOI-1257 - An unconfirmed planetary candidate (Santerne et al. 2014) with a very high eccentricity ($e_1 \sim 0.7$) that could lead to instabilities. This system is located in the borderline between the blue and the green regions, indicating that a second-order secular model is probably necessary to model its dynamical evolution;

\item HD 41004 Ab and Bb - A multi-planetary system. Planet $Bb$ is located very close to the $B$ star, with the dynamics properly described by the first-order model. Planet $Ab$, however, is located in the borderline region between the blue and green regions, which means that depending on the accuracy desired for the study, a second order approach may be necessary;

\item $\gamma$ Ceph Ab - Located in the boundary between the blue and green regions, again indicating that a second-order approach may be necessary. This system also presents a low osculating eccentricity ($e_1 = 0.05$), very close to the forced value $e_{1F}$. Due to the large perturbations from the binary companion, the origin of this planet has been subject of many studies (Th{\'e}bault et al. 2004; Giuppone et al. 2011, among others). A second-order secular theory has been proved necessary in analytical studies;

\item HD 196885 Ab - Located in the green region, this is another system for which a second-order model is necessary. The planet is located close to the MMR region, and perhaps high-order resonances could play should be taken into consideration. The dynamical evolution of the system and constraints to the orbital parameters have been focus of several works (Giuppone et al. 2012; Satyal et al. 2014);

\item $\alpha$ Cen Bb - Even though its existence is still under debate, the planetary candidate is located in the blue region and a first-order model is adequate to describe its secular dynamics. However, studies of the planetary formation around the B star, such as Th\'ebault et al. (2009), suggests a non-linearity of the eccentricity of putative planetesimals located at larger values of $a_1$, whose dynamics would require a second-order model;

\item Gl 86 Ab - Due to difficulties of determining the orbital parameters of the binary (Farihi et al. 2013), this system possesses a high indetermination of the semimajor axis and eccentricity of the binary. Nevertheless, in both cases the system is properly described by a first-order secular theory;

\item HD 126614 Ab - Located in the boundary between the blue and green regions, once again indicating the necessity of a second-order approach. Also, this system is very closely located to the MMR region, which would means that high-order resonances could play an important role in its dynamical evolution.

\end{itemize}

\section{Summary and Conclusions}
\label{discussions}

In this work we showed the importance and influence of high-order averaging theories in the study of S-type planetary orbits around tight binaries. To be assured that any difference that could arise with respect to simpler models would be only due to the averaging method itself, the disturbing function was expanded to high orders in semimajor axis ratio and eccentricities guaranteeing a precise representation for $a_1/a_2 < 0.4$ and $e_1,e_2 < 0.5$. We then used this expansion for the construction of a second-order secular model applying a Lie-series canonical perturbation technique. 

The basic properties of the secular dynamics are characterized by two quantities: the forced eccentricity $e_F$ and the secular frequency $g_S$. We defined and applied a geometric method (Section \ref{sec_ef}) to determine these quantities from a general secular Hamiltonian function. We showed that these can also be accurately obtained from N-body simulations with the aid of an iterative algorithm based on a quasi-periodic decomposition given by a frequency analysis method.

To compare the families of stationary solutions obtained from the secular models to those obtained from the N-body code, we introduced characteristic limits that define the applicability domain of each analytical theory. We calculated these limits to a large grid of parameters and constructed parametric planes that show, for any given system, whether it should be studied with a first, a second or higher-order model. These parametric planes also yield information concerning its orbital stability, the influence of MMRs on its dynamics and the magnitude of short-period oscillations.
We then applied these parametric planes to several real examples, including confirmed, candidates and contested planets in binary star systems. 

These planes show that there is always a region in the space of parameters that can be properly described by the first-order model for $0.1 \leq m_2/m_0 \leq 10$ and $e_2 \leq 0.6$. We also conclude that the second-order model is adequate up to higher values of $a_1/a_2$, but there is still a region that can not be properly described, specially for lower values of $e_2$ and larger values of  $ m_2/m_0$. We believe this region (white area in the parametric planes) should be properly described with a third or higher-order models. However, in many other cases, the limit of applicability of the second-order model coincides with the limit of orbital stability, indicating that higher-order models are unnecessary and will not improve the existing results.

\begin{acknowledgements}
Part of this work was developed during a visit of E.A-I. to the Universidad Nacional de Cordoba. We wish to express our gratitude to FAPESP (grants 2010/01209-2 and 2013/17102-0), CNPq, CONICET and Secyt/UNC for their support.
\end{acknowledgements}

\appendix

\section{Numerical Coefficients of the Disturbing Function}
\label{coef_num}

In this appendix we present the values of the coefficients $T_i$, $n_i$, $c_i$, $d_i$, $p^{(1)}_i$, $p^{(2)}_i$ and $p^{(3)}_i$ of the disturbing function (Eq. \ref{perturbadora11}). Two files containing these coefficients are provided as Electronic Supplementary Material of the journal and can be requested to the first author.

The file ``{\tt table-coefficients-ni\_ci\_di\_xi\_yi\_zi\_Ti.dat}'' contains the data of the development used throughout this paper. The file lines were ordered with respect to decreasing values of $|T_{i}\alpha^{n_i} e_1^{c_i} e_2^{d_i}|$ for $\alpha = 0.4$ and $e_1 = e_2 = 0.5$ and then arranged as described in Section \ref{angle-action}, with $N_0 = 72$, $N_S = 184$ and $N=10^4$. Using this development, the Second-order term of the Secular Hamiltonian (Eq. \ref{secondorder28}) has the order of $10^6$ terms. Table \ref{exemplo_coeficientes} shows an excerpt of the file as an example.

\begin{table}[h!]
\caption{Example of a section of the file ``{\tt table-coefficients-ni\_ci\_di\_xi\_yi\_zi\_Ti.dat}''. The file has 7 columns (from $n_i$ to $T_i$), with the first 6 columns composed of integers and the last column composed of real numbers. The index $i$ is the line of the file. Note that the Keplerian part ($i=-1,0$) of the Hamiltonian is not presented in this file.}
\label{exemplo_coeficientes}
\begin{tabular}{llllllll}
\hline\noalign{\smallskip}
File line ($i$)& $n_i$ & $c_i$ & $d_i$ & $p^{(1)}_i$ & $p^{(2)}_i$ & $p^{(3)}_i$ & $T_i$ \\
\noalign{\smallskip}\hline\noalign{\smallskip}
$ 1 $ & $ 2 $ & $ 0 $ & $ 0 $ & $ 0 $ & $ 0 $ & $ 0 $ & $ 0.25 $ \\
$ 2 $ & $ 2 $ & $ 0 $ & $ 2 $ & $ 0 $ & $ 0 $ & $ 0 $ & $ 0.375 $ \\
$ 3 $ & $ 2 $ & $ 2 $ & $ 0 $ & $ 0 $ & $ 0 $ & $ 0 $ & $ 0.375 $ \\
$ 4 $ & $ 2 $ & $ 2 $ & $ 2 $ & $ 0 $ & $ 0 $ & $ 0 $ & $ 0.5625 $ \\
$ 5 $ & $ 4 $ & $ 2 $ & $ 2 $ & $ 0 $ & $ 0 $ & $ 0 $ & $ 3.515625 $ \\
$ 6 $ & $ 2 $ & $ 0 $ & $ 4 $ & $ 0 $ & $ 0 $ & $ 0 $ & $ 0.46875 $ \\
$ 7 $ & $ 4 $ & $ 0 $ & $ 2 $ & $ 0 $ & $ 0 $ & $ 0 $ & $ 0.703125 $ \\
$ 8 $ & $ 4 $ & $ 2 $ & $ 0 $ & $ 0 $ & $ 0 $ & $ 0 $ & $ 0.703125 $ \\
$ 9 $ & $ 4 $ & $ 2 $ & $ 4 $ & $ 0 $ & $ 0 $ & $ 0 $ & $ 9.2285156 $ \\
$ 10 $ & $ 4 $ & $ 0 $ & $ 0 $ & $ 0 $ & $ 0 $ & $ 0 $ & $ 0.140625 $ \\
\noalign{\smallskip}\hline\noalign{\smallskip}
$ 73 $ & $ 3 $ & $ 1 $ & $ 1 $ & $ 0 $ & $ 0 $ & $ 1 $ & $ -0.9375 $ \\
$ 74 $ & $ 3 $ & $ 1 $ & $ 3 $ & $ 0 $ & $ 0 $ & $ 1 $ & $ -2.34375 $ \\
$ 75 $ & $ 5 $ & $ 1 $ & $ 3 $ & $ 0 $ & $ 0 $ & $ 1 $ & $ -8.6132813 $ \\
$ 76 $ & $ 5 $ & $ 1 $ & $ 1 $ & $ 0 $ & $ 0 $ & $ 1 $ & $ -1.640625 $ \\
$ 77 $ & $ 5 $ & $ 1 $ & $ 5 $ & $ 0 $ & $ 0 $ & $ 1 $ & $ -25.8398438 $ \\
$ 78 $ & $ 3 $ & $ 1 $ & $ 5 $ & $ 0 $ & $ 0 $ & $ 1 $ & $ -4.1015625 $ \\
$ 79 $ & $ 5 $ & $ 3 $ & $ 3 $ & $ 0 $ & $ 0 $ & $ 1 $ & $ -21.5332031 $ \\
$ 80 $ & $ 3 $ & $ 3 $ & $ 1 $ & $ 0 $ & $ 0 $ & $ 1 $ & $ -0.703125 $ \\
$ 81 $ & $ 5 $ & $ 3 $ & $ 1 $ & $ 0 $ & $ 0 $ & $ 1 $ & $ -4.1015625 $ \\
$ 82 $ & $ 5 $ & $ 3 $ & $ 5 $ & $ 0 $ & $ 0 $ & $ 1 $ & $ -64.5996094 $ \\
\noalign{\smallskip}\hline\noalign{\smallskip}
$ 185 $ & $ 6 $ & $ 3 $ & $ 6 $ & $ 5 $ & $ -10 $ & $ 6 $ & $ -2.9560257\times 10^{5} $ \\
$ 186 $ & $ 6 $ & $ 4 $ & $ 6 $ & $ 4 $ & $ -10 $ & $ 6 $ & $ 5.8235717\times 10^{5} $ \\
$ 187 $ & $ 6 $ & $ 3 $ & $ 8 $ & $ 5 $ & $ -10 $ & $ 6 $ & $ 1.1427852\times 10^{6} $ \\
$ 188 $ & $ 6 $ & $ 4 $ & $ 8 $ & $ 4 $ & $ -10 $ & $ 6 $ & $ -2.2513646\times 10^{6} $ \\
$ 189 $ & $ 6 $ & $ 4 $ & $ 6 $ & $ 6 $ & $ -10 $ & $ 6 $ & $ -5.4716639\times 10^{5} $ \\
$ 190 $ & $ 6 $ & $ 4 $ & $ 8 $ & $ 6 $ & $ -10 $ & $ 6 $ & $ 2.1153188\times 10^{6} $ \\
$ 191 $ & $ 6 $ & $ 5 $ & $ 6 $ & $ 5 $ & $ -10 $ & $ 6 $ & $ 1.0313413\times 10^{6} $ \\
$ 192 $ & $ 6 $ & $ 5 $ & $ 8 $ & $ 5 $ & $ -10 $ & $ 6 $ & $ -3.9871155\times 10^{6} $ \\
$ 193 $ & $ 5 $ & $ 3 $ & $ 7 $ & $ 4 $ & $ -10 $ & $ 5 $ & $ -1.9900952\times 10^{5} $ \\
$ 194 $ & $ 6 $ & $ 3 $ & $ 6 $ & $ 3 $ & $ -10 $ & $ 6 $ & $ 2.3447115\times 10^{5} $ \\
\noalign{\smallskip}\hline
\end{tabular}
\end{table}

The file ``{\tt table-coefficients-ni\_ci\_di\_xi\_yi\_zi\_Ti-short.dat}'' contains the data of a second development, valid for $\alpha < 0.3$, $e_1<0.1$ and $e_2 < 0.5$. The file is arranged again as described in Section \ref{angle-action}, with $N_0 = 72$, $N_S = 184$ and $N=1800$. Using this development, the Second-order term of the Secular Hamiltonian (Eq. \ref{secondorder28}) has the order of $5\times 10^3$ terms, which decreases substantially the computation time in comparison with the first file. Although it is a more limited development, it is still capable of reproducing the results presented in Section \ref{applications}, for orbits close to the secular stationary solution.

\section{Analytical Expressions for the Frequencies}
\label{ap01}

The frequencies $\nu_i$ on the left hand side of Eq. (\ref{secondorder13}) are given by

\begin{equation}
\begin{array}{rl}
\vspace{0.4cm}
\nu_1 = \displaystyle\frac{\partial {\cal H}_0}{\partial L_1^*} = & - {\cal G}^2 \displaystyle\sum_{l=-1}^{N_0} { T}_l {\cal M}_{n_l} {\cal Q}_{n_l} L_1^{{\rm a}_l} L_2^{{\rm b}_l}e_1^{{\rm c}_l} e_2^{{\rm d}_l}
\biggr( \displaystyle\frac{{\rm a}_l - {\rm c}_l}{L_1^*} + \displaystyle\frac{{\rm c}_l}{L_1^*e_1^{*2}} \biggl), \\
\vspace{0.4cm}
\nu_2 = \displaystyle\frac{\partial {\cal H}_0}{\partial L_2^*} = & - {\cal G}^2 \displaystyle\sum_{l=-1}^{N_0} { T}_l {\cal M}_{n_l} {\cal Q}_{n_l} L_1^{{\rm a}_l} L_2^{{\rm b}_l}e_1^{{\rm c}_l} e_2^{{\rm d}_l} 
\biggr( \displaystyle\frac{{\rm b}_l - {\rm d}_l}{L_2^*} + \displaystyle\frac{{\rm d}_l}{L_2^*e_2^{*2}} \biggl), \\
\vspace{0.4cm}
\nu_3 = \displaystyle\frac{\partial {\cal H}_0}{\partial G_1^*}= &  - {\cal G}^2 \displaystyle\sum_{l=-1}^{N_0} { T}_l {\cal M}_{n_l} {\cal Q}_{n_l} L_1^{{\rm a}_l} L_2^{{\rm b}_l}e_1^{{\rm c}_l} e_2^{{\rm d}_l} 
\biggr( \displaystyle\frac{{\rm d}_l\sqrt{1-e_2^{*2}}}{e_2^{*2}L_2^*} - \displaystyle\frac{{\rm c}_l\sqrt{1-e_1^{*2}}}{e_1^{*2}L_1^*} \biggl). \\
\end{array}
\label{secondorder16}
\end{equation}

The secular frequency for the Second-order Hamiltonian is calculated as Eq. (\ref{gsec1}) and gives

\begin{equation}
\begin{array}{rl}
\vspace{0.4cm}
g_{s2A}= & \nu_3 + \displaystyle\frac{{\cal G}^4}{4}\Biggl\{  \displaystyle\sum_{i=N_0+1}^{N} \  \displaystyle\sum_{j=N_S+1}^{N} \displaystyle\frac{\sigma_i \delta_{p^{(1)}_j,-p^{(1)}_i} \delta_{p^{(2)}_j,-p^{(2)}_i} \delta_{p^{(3)}_j,-p^{(3)}_i} T_i {\cal M}_{n_i} {\cal K}_{n_i}T_j {\cal M}_{n_j} {\cal K}_{n_j}}{-p^{(1)}_i\nu_1 - p^{(2)}_i\nu_2 + p^{(3)}_j\nu_3} \\
\vspace{0.4cm}
& \ \ \ \ \ \ \ \ \ \ \ \ \ \ L_1^{{\rm a}_i + {\rm a}_j} L_2^{{\rm b}_i+ {\rm b}_j}e_1^{{\rm c}_i+{\rm c}_j} e_2^{{\rm d}_i+{\rm d}_j} {\cal C}_{ij} \Biggr\} \\ 
\vspace{0.4cm}
& + \ \displaystyle\frac{{\cal G}^4}{4}\Biggl\{  \displaystyle\sum_{i=N_0+1}^{N} \  \displaystyle\sum_{j=N_S+1}^{N}  \displaystyle\frac{\sigma_i \delta_{p^{(1)}_j,-p^{(1)}_i} \delta_{p^{(2)}_j,-p^{(2)}_i} \delta_{p^{(3)}_j,-p^{(3)}_i}T_i {\cal M}_{n_i} {\cal K}_{n_i}T_j {\cal M}_{n_j} {\cal K}_{n_j}}{p^{(1)}_i\nu_1 + p^{(2)}_i\nu_2 + p^{(3)}_j\nu_3} \\
\vspace{0.4cm}
&  \ \ \ \ \ \ \ \ \ \ \ \ \ \ L_1^{{\rm a}_i + {\rm a}_j} L_2^{{\rm b}_i+ {\rm b}_j}  e_1^{{\rm c}_i+{\rm c}_j} e_2^{{\rm d}_i+{\rm d}_j} {\cal D}_{ij} \Biggr\}, \\ 
\end{array}
\label{gs2ndan}
\end{equation}

\noindent where

\begin{equation}
\begin{array}{rl}
\vspace{0.4cm}
{\cal C}_{ij} = & \Biggl[p^{(1)}_i \biggl(\displaystyle\frac{{\rm a}_j + {\rm a}_i - {\rm c}_j- {\rm c}_i}{L_1^*} + \displaystyle\frac{{\rm c}_j+{\rm c}_i}{L_1^*e_1^{*2}}\biggr) - p^{(2)}_i \biggl(\displaystyle\frac{{\rm b}_j + {\rm b}_i - {\rm d}_j - {\rm d}_i}{L_2^*} + \displaystyle\frac{{\rm d}_j + {\rm d}_i}{L_2^*e_2^{*2}} \biggr) \\ 
\vspace{0.4cm}
&+p^{(3)}_i\biggl(\displaystyle\frac{{\rm d}_j\sqrt{1-e_2^{*2}}}{L_2^*e_2^{*2}} - \displaystyle\frac{{\rm c}_j\sqrt{1-e_1^{*2}}}{L_1^*e_1^{*2}} \biggr) - \ p^{(3)}_j\biggl(\displaystyle\frac{{\rm d}_i\sqrt{1-e_2^{*2}}}{L_2^*e_2^{*2}} - \displaystyle\frac{{\rm c}_i\sqrt{1-e_1^{*2}}}{L_1^*e_1^{*2}} \biggr) \Biggr] \\ 
\vspace{0.4cm}
& \times \Biggl[ \displaystyle\frac{({\rm d}_i + {\rm d}_j) \sqrt{1-e_2^{*2}}}{L_2^* e_2^{*2}} -  \displaystyle\frac{({\rm c}_i + {\rm c}_j) \sqrt{1-e_1^{*2}}}{L_1^* e_1^{*2}} \Biggl] \\
\vspace{0.4cm}
& + \displaystyle\frac{1}{L_1^{*2} e_1^{*4}} \Biggl[ 2p^{(1)}_i({\rm c}_i + {\rm c}_j)\sqrt{1 - e_1^{*2}}  + (1 - e_1^{*2}) (p^{(3)}_j {\rm c}_i - p^{(3)}_i {\rm c}_j)\Biggl] \\
\vspace{0.4cm}
& - \displaystyle\frac{1}{L_2^{*2} e_2^{*4}} \Biggl[ 2p^{(2)}_i({\rm d}_i + {\rm d}_j)\sqrt{1 - e_2^{*2}}  + (1 - e_2^{*2})(p^{(3)}_i{\rm d}_j - p^{(3)}_j {\rm d}_i) \Biggl], \\
\end{array}
\label{secondorder33}
\end{equation}

\begin{equation}
\begin{array}{rl}
\vspace{0.4cm}
{\cal D}_{ij} = & - \Biggl[ p^{(1)}_i \biggl(\displaystyle\frac{{\rm a}_j-{\rm c}_j - {\rm a}_i + {\rm c}_j}{L_1^*} + \displaystyle\frac{{\rm c}_j - {\rm c}_i}{L_1^*e_1^{*2}}\biggr) + \  p^{(2)}_i \biggl(\displaystyle\frac{{\rm b}_j-{\rm d}_j - {\rm b}_i + {\rm d}_j}{L_2^*} + \displaystyle\frac{{\rm d}_j - {\rm d}_i}{L_2^*e_2^{*2}}\biggr) \\ 
\vspace{0.4cm}
& + \ p^{(3)}_i\biggl(\displaystyle\frac{{\rm d}_j\sqrt{1-e_2^{*2}}}{L_2^*e_2^{*2}} - \displaystyle\frac{{\rm c}_j\sqrt{1-e_1^{*2}}}{L_1^*e_1^{*2}} \biggr) + \ p^{(3)}_j\biggl(\displaystyle\frac{{\rm d}_i\sqrt{1-e_2^{*2}}}{L_2^*e_2^{*2}} - \displaystyle\frac{{\rm c}_i\sqrt{1-e_1^{*2}}}{L_1^*e_1^{*2}} \biggr) \Biggr] \\ 
\vspace{0.4cm}
& \times \Biggl[ \displaystyle\frac{({\rm d}_i + {\rm d}_j) \sqrt{1-e_2^{*2}}}{L_2^* e_2^{*2}} -  \displaystyle\frac{({\rm c}_i + {\rm c}_j) \sqrt{1-e_1^{*2}}}{L_1^* e_1^{*2}} \Biggl] \\
\vspace{0.4cm}
& + \displaystyle\frac{1}{L_1^{*2} e_1^{*4}} \Biggl[ - 2 p^{(1)}_i({\rm c}_i + {\rm c}_j)\sqrt{1 - e_1^{*2}}  + (1 - e_1^{*2}) (p^{(3)}_j {\rm c}_i - p^{(3)}_i {\rm c}_j)\Biggl] \\
\vspace{0.4cm}
& + \displaystyle\frac{1}{L_2^{*2} e_2^{*4}} \Biggl[ 2 p^{(2)}_i({\rm d}_i + {\rm d}_j)\sqrt{1 - e_2^{*2}}  + (1 - e_2^{*2})(p^{(3)}_i{\rm d}_j - p^{(3)}_j {\rm d}_i) \Biggl]. \\
\end{array}
\label{secondorder34}
\end{equation}

\end{document}